\documentclass[preprint,tightenlines,showpacs,amsmath,amssymb]{revtex4-1}

\usepackage[english]{babel}
\usepackage{afterpage}
\usepackage{graphicx}
\usepackage{dcolumn}
\usepackage{bm}
\usepackage{color}
\usepackage{slashed}
\usepackage{latexsym}

\DeclareMathAlphabet\mathbfcal{OMS}{cmsy}{b}{n}

\newcommand{\dd}{\mathrm{d}}  
\newcommand{\ii}{\mathrm{i}}  
\newcommand{\ee}{\mathrm{e}}  

\newcommand{\op}[1]{\hat{#1}} 
\newcommand{\bra}[1]{\langle{#1}\vert} 
\newcommand{\ket}[1]{\vert{#1}\rangle} 
\newcommand{\scal}[2]{\langle #1 | #2 \rangle} 
\newcommand{\sscal}[2]{\langle\!\langle #1 | #2 \rangle\!\rangle} 

\begin{document}

\title{Unitary vs pseudo-unitary time evolution and statistical effects in the dynamical Sauter-Schwinger process}
\author{K. Krajewska}
\author{J. Z. Kami\'nski}

\affiliation{$^1$Institute of Theoretical Physics, Faculty of Physics, University of Warsaw, Pasteura 5,
02-093 Warszawa, Poland}
\date{\today}

\begin{abstract}
Dynamical Sauter-Schwinger mechanism of pair creation by a time-dependent electric field comprising of $N_{\rm rep}$ identical pulses is analyzed within the framework of the spinor and scalar quantum electrodynamics.
For linearly polarized pulses, both theories predict that a single eigenmode of the matter wave follows the dynamics of a two-level system. This dynamics, however,
is either governed by a Hermitian (for spin 1/2 particles) or pseudo-Hermitian (for spin 0 particles) Hamiltonian. Essentially, both theories lead to a Fraunhofer-type 
enhancement of the momentum distributions of created pairs. While in the fermionic case the enhancement is never perfect and it deteriorates with increasing the number of
pulses in a train $N_{\rm rep}$, in the bosonic case we observe the opposite. More specifically, it is at exceptional points where the spectra of bosonic pairs scale exactly as $N_{\rm rep}^2$, and this scaling
is even enhanced with increasing the number of pulses in a train.
\end{abstract}

\maketitle

\section{Introduction}
\label{sec::intro}

Diffraction and interference of waves~\cite{Crawford} have played the fundamental role in the development of science. While both phenomena have been observed for 
sound~\cite{Rayleigh} and surface waves~\cite{Whitham}, it is the diffraction and interference of light discovered by F. M. Grimaldi (1665) and T. Young (1800)
(see, Refs.~\cite{BornWolf,Fremont} for details), that has paved the way to development of modern physics. 
Both phenomena are present, for instance, in light scattering by a diffraction grating. Specifically, in the far field zone, the resulting intensity 
of the monochromatic light scattered by $N_{\rm rep}$ parallel slits can be described by the Fraunhofer formula~\cite{Crawford},
\begin{equation}
I(u)=I_0D(u)\Bigl(\frac{\sin(N_{\rm rep}\pi a u)}{\sin(\pi a u)}\Bigr)^2.
\label{intro1}
\end{equation}
Here, $u=\sin\theta/\lambda$ is related to the scattering angle $\theta$ and the wavelength of the incident wave $\lambda$, $a$ is the distance between two subsequent slits, whereas $I_0$ is the incident light intensity. 
The Fraunhofer formula essentially consists of two factors. One of them, $\bigl(\frac{\sin(N_{\rm rep}\pi a u)}{\sin(\pi a u)}\bigr)^2$, is called the {\it interference term}.
It is responsible for the coherent $N_{\rm rep}^2$-type enhancement of the scattered wave if detected at the angle $\theta$ such that $a u$ is integer. The factor $D(u)$, on the other hand, is 
called the {\it diffraction factor}. It describes the wave scattering off a single slit. It depends only on the shape of the individual slit, 
provided that the neighboring slits are sufficiently well separated from each other. In most cases $D(u)$ is a slowly varying function of $u$, as opposed to the rapidly 
changing interference term. For this reason the general pattern~\eqref{intro1} consists of well-separated and narrow interference peaks, the intensities of which are modulated by the diffraction term. 
It appears, however, that in some cases (for instance, when the linear size of the slits becomes comparable to their separation) the diffraction term also exhibits sharp peaks. 
In such circumstances the distinction between the diffraction and interference peaks is rather difficult to trace. This may lead to misinterpretation 
of some features of the pattern~\eqref{intro1}, as we shall discuss in our paper.

With the emergence of quantum theories and the discovery of wave properties of matter, the investigation of diffraction and interference phenomena of matter waves became very important 
from the fundamental as well as the practical point of views. The point being that these phenomena have prompted some unexpected observations and applications 
(see, e.g.,~\cite{VanHove,Silverman1995,Silverman2008,Deymier}), such as in the low-energy~\cite{Peierls1979} or high-energy~\cite{BaronePredazzi} scattering. 
Another example is the so-called diffraction radiation~\cite{Potylitsyn}, which is emitted when charged particles move in vacuum along a periodically deformed surface;
the latter playing the role of a diffraction grating. This is known as the Smith-Purcell effect~\cite{SmithPurcell} and it can be applied, for instance, for the generation 
of teraherz radiation, which finds considerable interest in physics, chemistry, and biology~\cite{Williams2006}. Closely related to the Smith-Purcell effect is the generation 
of coherent frequency combs of radiation in the scenario in which electrons (or other charged particles) interact with a train of strong laser pulses. In such case the pulse train acts as
a diffraction grating in the time domain~\cite{K1,F1}. Note that the coherent frequency combs of radiation generated from Compton or Thomson scattering offer a possibility for the diagnosis 
of relativistically intense and short laser pulses~\cite{K2}. Moreover, similar combs have been observed for matter waves. Specifically, the multislit interference and diffraction
pattern, as the one predicted by Eq.~\eqref{intro1}, has been observed in the momentum and energy distributions of particles emitted via
the Breit-Wheeler electron-positron pair creation~\cite{K3} or in photoionization~\cite{K5,K6}. 
These selected examples show that the Fraunhofer formula~\eqref{intro1} is universal, as it can be applied across different areas of classical and quantum physics.

The aim of this paper is to investigate the quantum vacuum instabilities caused by the action of time-dependent electric fields; the process known as the dynamical Sauter-Schwinger 
pair creation. In this context, it has been demonstrated that the multislit interference and diffraction pattern in the momentum distribution of created particles is observed
when a finite sequence of electric field pulses interacts with the vacuum~\cite{Akkermans,Li1,Li2,KTK,KKproc}. Here, we generalize our recent results~\cite{KTK,KKproc}
by comparing theoretical approaches toward particle-antiparticle pair creation based on either the spinor quantum
electrodynamics (QED) or scalar QED. In other words, we are interested in investigating the effect of statistics on the Fraunhofer-type enhancements in pair production. 
In this context, it is important to mention the paper by Li {\it et al.}~\cite{Li1} where such effects were already studied.
This was done by solving the quantum Vlasov equation~\cite{Schmidt} (see, in Appendix~\ref{Vlasov}). The main conclusion of~\cite{Li1} was that,
while the momentum distributions of scalar and spinor particles exhibit very similar Fraunhofer peak patterns, they are shifted relative to each other. Such shifting was ascribed
to different statistics of produced particles. However, as we will show, for the parameters considered in our paper this is not necessarily the case. 
Instead, we shall focus on a fundamental difference between both theories, which is the unitary versus pseudo-unitary time evolution of the respective fermionic and bosonic fields.
Its consequences on the resulting Fraunhofer-like enhancements described above will be studied in this paper in great detail.

Note that our work fits nicely in the upgrowing area of research devoted to non-Hermitian quantum theories.
The reason being that in the case of scalar pair production, the dynamics of a single eigenmode of the bosonic field is determined by a pseudo-Hermitian Hamiltonian
[Eq.~\eqref{bf29}]. Thus, our work adds to a long list of potential applications of pseudo-Hermitian theories that includes generalized coherent states~\cite{Wodkiewicz}, 
synthetic optical lattices~\cite{Makris,Guo}, waveguide couplers~\cite{Ruter,Lee}, 
laser cavities~\cite{Peng,Feng}, or Rabi systems~\cite{Torosov,Kus} (for more application, see, also~\cite{Rotter1,Moiseyev,Rotter2}). Interestingly, in this context the role of the so-called
{\it exceptional points} is frequently studied~\cite{Peng,Feng,Praxmeyer}. While non-Hermitian Hamiltonians have complex eigenvalues, 
at those points their eigenvalues coalesce. In other words, they exhibit a {\it non-avoided crossing} where their real components are identical, as are their
imaginary ones. This leads to counterintuitive effects when steering the system in the vicinity of the exceptional points (see, for instance, Refs.~\cite{Peng,Feng}).
As we show in this paper, the exceptional points are also found in the dynamical pair production of spin 0 particles, and it is only at those points that
the fully coherent enhancement of the respective particle spectra is observed. Note that our problem relies on studying a two-state dynamics.
Therefore, the conclusions drawn from our results apply essentially to any system such that its dynamics can be traced back to that of a two-level system.

The paper is organized as follows. In Sec.~\ref{theory}, we shall present the theoretical formulation of the dynamical Sauter-Schwinger process using either the scalar or spinor QED. 
Momentum distributions of created pairs based on both these theories will be presented in Sec.~\ref{results}. Also in Sec.~\ref{results}, we will provide an analytical explanation of our numerical 
results arising from the analysis of the operators that evolve in time bosonic and fermionic fields. The properties
of those operators will be analyzed in Appendix~\ref{SU(1,1)} and~\ref{appendix1}, respectively. In Sec.~\ref{conclusions}, we will summarize our results.

The numerical results will be expressed in relativistic units. Specifically,
we shall use the Sauter-Schwinger electric field strength $\mathcal{E}_S=m_{\mathrm{e}}^2c^3/|e|\hbar$, with the corresponding
strength of the vector potential, $A_S=m_{\mathrm{e}}c/|e|$, as well as the Compton time $t_C=\hbar/m_{\mathrm{e}}c^2$.
Here, $m_{\rm e}$ is the electron mass and $e=-|e|<0$ is its charge.
Since now on, in our theoretical formulation we shall keep $\hbar=1$ and an arbitrary mass of created particles $m$. However, we will choose $m=m_{\rm e}=c=\hbar=1$ in our numerical calculations.

\section{Theoretical formulation}
\label{theory}

The spontaneous formation of particle-antiparticle pairs by a homogeneous in space, time-dependent electric field is studied 
in this paper within the scalar and the spinor QED frameworks. In order to elucidate the differences between both approaches, we 
present below both theoretical formulations. Typically, such comparison has been performed within the quantum kinetic approach. 
Thus, concealing very subtle but fundamental features of quantum dynamics. Here, we extend our previous studies~\cite{KTK,KKproc}
to scalar QED. As we show, while the spinor QED facilitates a typical unitary time-evolution of the respective fermionic field
eigenmodes~\cite{KTK,KKproc}, the respective time evolution of the bosonic field eigenmodes is pseudo-unitary.

\subsection{Electric field description}
\label{electric}
\begin{figure}
\includegraphics[width=0.9\textwidth]{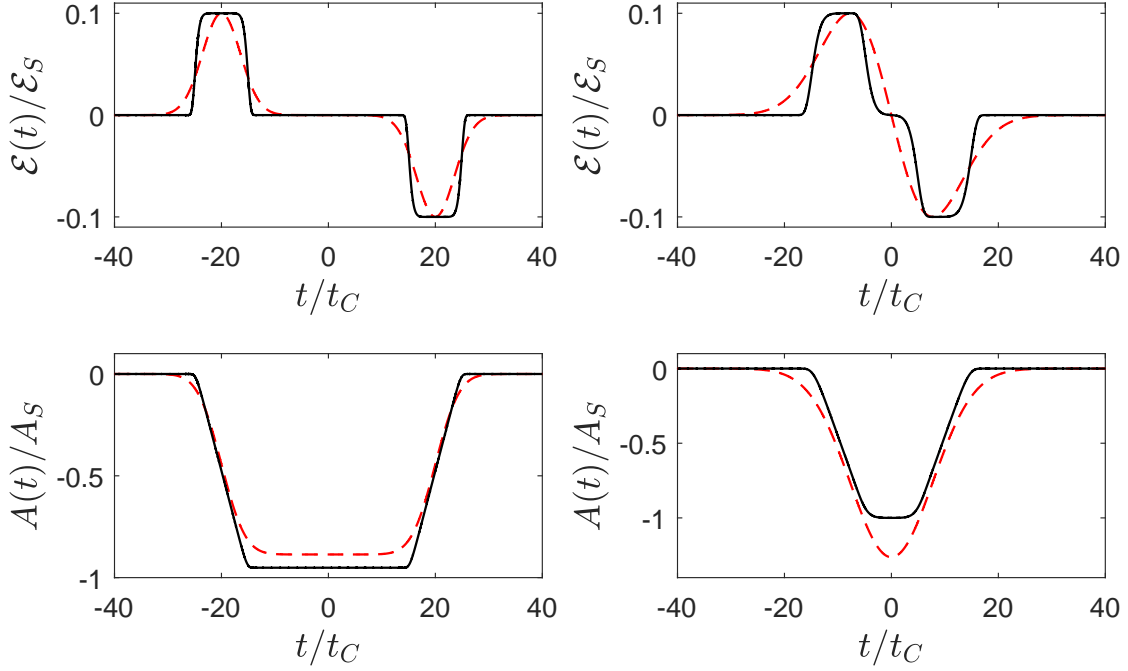}
\caption{Time-dependence of the electric field $\mathcal{E}(t)$ and the vector potential $A(t)$ (in relativistic units) 
for a single pulse considered in this paper ($N_{\rm rep}=1$). The shape functions in~\eqref{a8} are for $\sigma=5/t_C$, $T_0=40t_C$ 
(left column), and for $\sigma=10/t_C$, $T_0=10t_C$ (right column). In both cases, the amplitude of the electric field
is $\mathcal{E}_0=-0.1\mathcal{E}_S$. Also, the dashed lines are for $M=1$, whereas the solid lines are for $M=5$. As one can see, with increasing $M$,
the shape of super-Gaussian envelopes becomes similar to the rectangular one. The qualitative difference between both columns is that, while in the left column
both half-pulses are well separated in time, in the right column they overlap. In other words, the single pulse in the left column is much longer that the one
on the right. As we will show later, this will strongly affect the structure of diffraction patterns observed in the momentum distributions of created pairs.
\label{dirackleinpolav2}}
\end{figure}

We consider an electric field which oscillates linearly along the $z$-direction, ${\mathbfcal{E}}(t)={\cal E}(t){\bm e}_z$. In addition, we assume that it satisfies the condition,
\begin{equation}
\int_{-\infty}^{+\infty}\dd t\,{\cal E}(t)=0.
\label{a2}
\end{equation}
If the electric field is defined as
\begin{equation}
{\cal E}(t)={\cal E}_0F(t),
\label{a3}
\end{equation}
where ${\cal E}_0$ is the field amplitude whereas $F(t)$ is its shape function, then it follows from~\eqref{a2} that
\begin{equation}
\int_{-\infty}^{+\infty}\dd t\,F(t)=0.
\label{a4}
\end{equation}
The significance of Eqs.~\eqref{a2} and~\eqref{a4} becomes clear when we introduce the vector potential; namely,
${\bm A}(t)=A(t){\bm e}_z$ where ${\cal E}(t)=-{\rm d}A(t)/{\rm d}t$. In other words, if
\begin{equation}
A(t)={\cal E}_0f(t),
\label{a5}
\end{equation}
then
\begin{equation}
f(t)=f(+\infty)+\int_t^{+\infty}\dd t'F(t').
\label{a6}
\end{equation}
Here, taking into account Eq.~\eqref{a4}, we conclude that $f(-\infty)=f(+\infty)$. This means that in the remote past and in the far future,
\begin{equation}
\underset{t\rightarrow-\infty}{\lim}A(t)=\underset{t\rightarrow+\infty}{\lim}A(t).
\label{a1}
\end{equation}
Without loosing the generality, we can put this constant to zero and, equivalently, $f(-\infty)=f(+\infty)=0$. In such case, the behavior of the vector potential $A(t)$ 
guarantees that our asymptotic ``in'' and ``out'' states will be indeed field-free states. This is particularly important for the Sauter-Schwinger pair creation and, hence, 
it justifies imposing the condition~\eqref{a2}.

In the following, we shall consider a single pulse with the shape function in~\eqref{a3} given by
\begin{equation}
F_1(t)=\exp\Bigl[-\Bigl(\frac{t-T_0/2}{\sigma}\Bigr)^{2M}\Bigr]-\exp\Bigl[-\Bigl(\frac{t+T_0/2}{\sigma}\Bigr)^{2M}\Bigr].
\label{a8}
\end{equation}
Note that each half-pulse is given by either Gaussian ($M=1$) or super-Gaussian ($M>1$) envelope, with a bandwidth $\sigma$ and a time delay between them $T_0$.
An interesting property of super-Gaussian envelopes is that, while they remain smooth functions (of class $C^{\infty}(\mathbb{R})$), they approach the step function,
\begin{equation}
S(t)=\begin{cases} 1, & t\in [-\sigma,\sigma], \cr 0 , & t\notin (-\sigma,\sigma),\end{cases}
\label{a9}
\end{equation}
for large $M$. This is illustrated in the upper panel of Fig.~\ref{dirackleinpolav2}. In the lower panel, we present the time-dependence of the corresponding
vector potential~\eqref{a5}. The difference between both columns in Fig.~\ref{dirackleinpolav2} is the time delay $T_0$ between both half-pulses.

\begin{figure}
\includegraphics[width=0.9\textwidth]{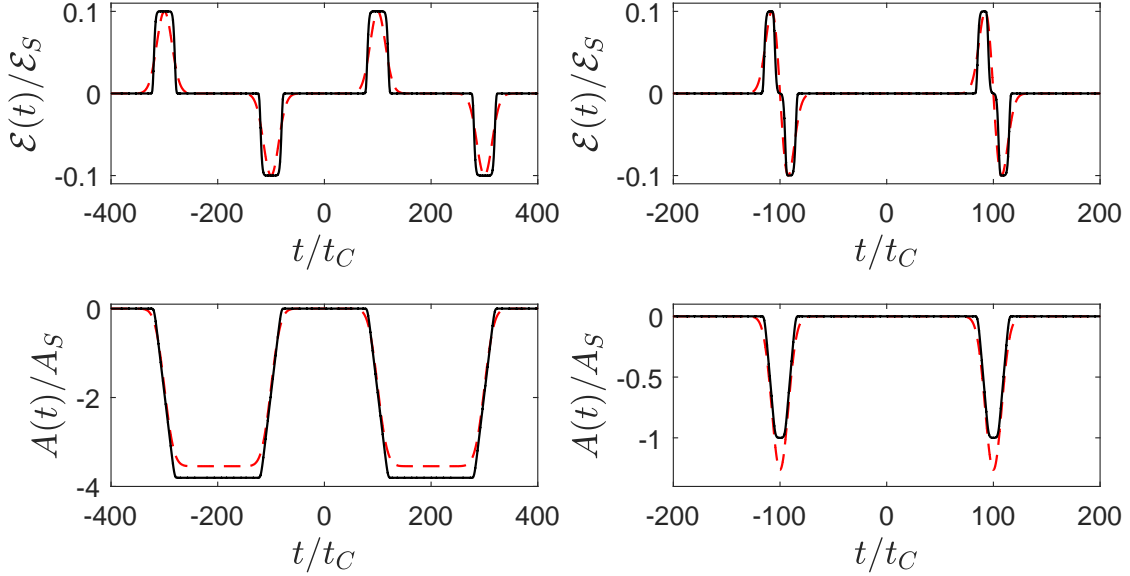}
\caption{The same as in Fig.~\ref{dirackleinpolav2} but for a sequence of two pulses [$N_{\rm rep}=2$ in Eq.~\eqref{a10}]. In addition, we have
$\sigma=20/t_C$, $T_0=200t_C$, $T=400t_C$ (left column), and $\sigma=10/t_C$, $T_0=10t_C$, $T=200t_C$ (right column).  
The remaining parameters of the field are the same as in Fig.~\ref{dirackleinpolav2}.
\label{dirackleinpola}}
\end{figure}

Similarly, we shall also consider a train consisting of $N_{\rm rep}$ such pulses, with
\begin{equation}
F_{N_{\mathrm{rep}}}(t)=\mathcal{N}\sum_{\ell=1}^{N_{\mathrm{rep}}}F_1\Bigl[t+\Bigl(\ell-\frac{1}{2}(N_{\mathrm{rep}}+1)\Bigr)T\Bigr].
\label{a10}
\end{equation}
Here, the normalization constant $\mathcal{N}$ is chosen such that
\begin{equation}
\max_t |F_{N_{\mathrm{rep}}}(t)|=1,
\label{sg7}
\end{equation}
to make sure that the maximum intensity of the electric field is independent of the parameters chosen in the above definitions. In Fig.~\ref{dirackleinpola},
we represent the respective sequence of two pulses ($N_{\rm rep}=2$). Again, both columns in the figure are plotted for different values of $T_0$ (and $T$). This obviously has to 
influence the interference-diffraction pattern observed in the momentum distributions of created particles, the details of which will be presented in Sec.~\ref{numerical}.

This model of the oscillating in time electric field will be used in Sec.~\ref{numerical} to illustrate the general theory derived in the next two sections. As first, we shall consider the 
bosonic pair creation. Its rigorous treatment is based on the Klein-Gordon equation, which is the foundation of scalar QED.

\subsection{Scalar QED}
\label{BB}

We define the Klein-Gordon boson field operator $\hat{\Phi}(x)$ as
\begin{equation}
\hat{\Phi}(x)=\int\frac{\dd^3{\bm p}}{(2\pi)^3}\Bigl(\Phi_{\bm p}^{(+)}(x)\hat{b}_{\bm p}+\Phi_{\bm p}^{(-)}(x)\hat{d}_{-{\bm p}}^\dagger\Bigr),
\label{bf1}
\end{equation}
where $\Phi_{\bm p}^{(\pm)}(x)$ are the one-particle solutions of the Klein-Gordon equation, 
whereas $\op{b}_{\bm p}$ ($\op{d}_{\bm p}$) is the annihilation operator of a boson (antiboson) with a given momentum ${\bm p}$. These operators define the in-vacuum state
through the conditions: $\op{b}_{\bm p}\ket{0_{-\infty}}=0$ and $\op{d}_{\bm p}\ket{0_{-\infty}}=0$. They also satisfy the commutation relations,
\begin{equation}
[\op{b}_{\bm p},\op{b}_{{\bm p}'}^\dagger]=[\op{d}_{\bm p},\op{d}_{{\bm p}'}^\dagger]=\delta^{(3)}({\bm p}-{\bm p}'),
\label{bf2}
\end{equation}
with the remaining commutators equal to zero. $\Phi_{\bm p}^{\pm}(x)$, on the other hand, will be constructed in the next section.

\subsubsection{One-particle solutions of the Klein-Gordon equation}
\label{one-particle-bosons}

Our aim is to solve the scalar Klein-Gordon equation coupled to the external electromagnetic field,
\begin{equation}
\Bigl[\bigl(\ii\partial-eA\bigr)^2-(mc)^2\Bigr]\Phi(x)=0.
\label{kg1}
\end{equation}
Since the problem has translational symmetry, one can look for those solutions $\Phi(x)$ in the separable form,
\begin{equation}
\Phi(x)=\ee^{\ii {\bm p}\cdot {\bm x}}\Phi_{\bm p}(t),
\label{kg2}
\end{equation}
where ${\bm p}$ is the particle asymptotic momentum. Substituting~\eqref{kg2} into~\eqref{kg1}, 
we obtain that the function $\Phi_{\bm p}(t)$ satisfies the harmonic oscillator equation,
\begin{equation}
\Bigl[\frac{\dd^2}{\dd t^2}+\omega_{\bm p}^2(t)\Bigr]\Phi_{\bm p}(t)=0,
\label{kg3}
\end{equation}
 with the time-dependent frequency $\omega_{\bm p}(t)$,
\begin{equation}
\omega_{\bm p}(t)=\sqrt{(mc^2)^2+c^2{\bm p}_\perp^2+c^2(p_{\|}-eA(t))^2}.
\label{kg4}
\end{equation}
Here, we have introduced the longitudinal $p_{\|}$ and the transverse ${\bm p}_\perp$ components of the particle asymptotic momentum such that
\begin{equation}
p_{\|}={\bm p}\cdot{\bm e}_z,\quad {\bm p}_\perp={\bm p}-p_{\|}{\bm e}_z.
\label{kg5}
\end{equation}
To interpret the solutions of Eq.~\eqref{kg3}, we realize that in the remote past ($t\rightarrow -\infty$) this equation becomes
\begin{equation}
\Bigl[\frac{\dd^2}{\dd t^2}+\omega_{\bm p}^2\Bigr]\Phi_{\bm p}(t)=0,
\label{kg6}
\end{equation}
where 
\begin{equation}
\omega_{\bm p}=\sqrt{(mc^2)^2+c^2{\bm p}^2}.
\label{kg7}
\end{equation}
Therefore, there exist two linearly independent solutions of Eq.~\eqref{kg6} which we will label by the parameter $\beta$,
\begin{equation}
\Phi_{\bm p}^{(\beta)}(t)\underset{t\rightarrow -\infty}{\sim}\ee^{-\ii\beta\omega_{\bm p}t}.
\label{kg8}
\end{equation}
The one corresponding to a positive energy (with $\beta=+$) will be interpreted as a boson whereas the other one (with $\beta=-$) as
an antiboson. In this way, we have determined two sets of solutions of the Klein-Gordon equation that appear in~\eqref{bf1},
\begin{equation}
\Phi^{(\beta)}_{\bm p}(x)=\ee^{\ii {\bm p}\cdot {\bm x}}\Phi_{\bm p}^{(\beta)}(t),
\label{kg9}
\end{equation}
where $\Phi_{\bm p}^{(\beta)}(t)$ solves Eq.~\eqref{kg3} and asymptotically behaves according to~\eqref{kg8}. Note that we have chosen the same symbol for the space-time- and time-dependent
solutions, discriminating them by the argument $x=(ct,{\bm x})$ and $t$.

It is crucial that $\Phi^{(\beta)}_{\bm p}(x)$ form an orthonormal and complete set of solutions of the Klein-Gordon equation. This is provided that 
the inner product of two such wave functions, $\Phi_{\bm p}^{(\beta)}(x)$ and $\Phi_{{\bm p}'}^{(\beta')}(x)$, is defined as~\cite{Grib1,Grib2,Greiner1,Greiner2}
\begin{align}
\langle\Phi_{\bm p}^{(\beta)}|\Phi_{{\bm p}'}^{(\beta')}\rangle&=\ii\int\dd^3{\bm x}\,[\Phi_{\bm p}^{(\beta)}(x)]^*\overleftrightarrow{\partial_0}\Phi_{{\bm p}'}^{(\beta')}(x)\nonumber\\
&=\ii\int\dd^3{\bm x}\,\Bigl([\Phi_{\bm p}^{(\beta)}(x)]^*[\partial_0\Phi_{{\bm p}'}^{(\beta')}(x)]
-[\partial_0\Phi_{\bm p}^{(\beta)}(x)]^*\Phi_{{\bm p}'}^{(\beta')}(x)\Bigr).
\label{kg10}
\end{align}
Its physical significance can be realized when considering the four-current density of charge, $j^\mu(x)$. 
In the absence of the external electromagnetic field, it is defined as~\cite{Greiner1,Greiner2}
\begin{equation}
j^\mu(x)=\frac{\ii e}{2m}\Bigl(\Phi^{(\beta)*}_{\bm p}(x)\bigl(\partial^\mu\Phi^{(\beta)}_{\bm p}(x)\bigr)-\bigl(\partial^\mu\Phi^{(\beta)*}_{\bm p}(x)\bigr)
\Phi^{(\beta)}_{\bm p}(x)\Bigr),
\label{kg11}
\end{equation}
where $j^\mu(x)$ satisfies the continuity equation,
\begin{equation}
\partial_\mu j^\mu(x)=0.
\label{kg12}
\end{equation}
Taking into account~\eqref{kg9}, it follows from this equation that the quantity,
\begin{equation}
Q=\int\dd^3{\bm x}j^0(x),
\label{kg13}
\end{equation}
is conserved. Since $j^0(x)$ can take either positive or negative values, it cannot be identified with the probability density. Instead, 
if we reinterpret the Klein-Gordon equation as satisfied by a quantum field $\hat{\Phi}(x)$, $\op{j}^0(x)$ will describe the charge density of the field, 
whereas $\op{Q}$ will be the field charge~\cite{Greiner1,Greiner2}. Going back to the definition of the Klein-Gordon inner 
product~\eqref{kg10}, we see therefore that it is related to the conservation of the field charge $Q$. In the presence of the electromagnetic field, the four-vector current 
density has to be redefined,
\begin{align}
j^\mu(x)=\frac{\ii e}{2m}\Bigl(\Phi^{(\beta)*}_{\bm p}(x)\bigl(\partial^\mu\Phi^{(\beta)}_{\bm p}(x)\bigr)-\bigl(\partial^\mu\Phi^{(\beta)*}_{\bm p}(x)\bigr)
\Phi^{(\beta)}_{\bm p}(x)\Bigr)-\frac{e^2}{mc}A^\mu(x)|\Phi_{\bm p}^{(\beta)}(x)|^2.
\label{kg14}
\end{align}
Nevertheless, for as long as $A^0(x)=0$, which is the case discussed here, the definition of the Klein-Gordon inner product~\eqref{kg10}
does not change.

Keeping this in mind, we obtain for the inner product of the scalar wave functions~\eqref{kg10},
\begin{align}
\langle\Phi_{\bm p}^{(\beta)}|\Phi_{{\bm p}'}^{(\beta')}\rangle&=(2\pi)^3\delta^{(3)}({\bm p}-{\bm p}')\Bigl(\frac{\ii}{c}[\Phi_{\bm p}^{(\beta)}(t)]^*
\dot{\Phi}_{\bm p}^{(\beta')}(t)-\frac{\ii}{c}[\dot{\Phi}_{\bm p}^{(\beta)}(t)]^*\Phi_{\bm p}^{(\beta')}(t)\Bigr).\label{kg15}
\end{align}
Using Eqs.~\eqref{kg3} and~\eqref{kg15}, one can show that 
\begin{equation}
\frac{\dd}{\dd t}\Bigl(\frac{\ii}{c}[\Phi_{\bm p}^{(\beta)}(t)]^*
\dot{\Phi}_{\bm p}^{(\beta')}(t)-\frac{\ii}{c}[\dot{\Phi}_{\bm p}^{(\beta)}(t)]^*\Phi_{\bm p}^{(\beta')}(t)\Bigr)=0,
\label{kg16}
\end{equation}
where the dot stands for time derivative. It follows from this equation that the quantity in the brackets is constant in time. Setting its value at $t\rightarrow-\infty$ gives
\begin{align}
\underset{t\rightarrow-\infty}{\lim}\Bigl(\frac{\ii}{c}[\Phi_{\bm p}^{(\beta)}(t)]^*
\dot{\Phi}_{\bm p}^{(\beta')}(t)-\frac{\ii}{c}[\dot{\Phi}_{\bm p}^{(\beta)}(t)]^*\Phi_{\bm p}^{(\beta')}(t)\Bigr)
=\frac{2\omega_{\bm p}}{c}\beta\delta_{\beta\beta'}.
\label{kg17}
\end{align}
Hence, the one-particle solutions of the Klein-Gordon equation $\Phi_{\bm p}^{(\beta)}(x)$ can be normalized such that
\begin{equation}
\ii\int\dd^3{\bm x}\,[\Phi_{\bm p}^{(\beta)}(x)]^*\overleftrightarrow{\partial_0}\Phi_{{\bm p}'}^{(\beta')}(x)=2mc\beta(2\pi)^3\delta^{(3)}({\bm p}-{\bm p}')\delta_{\beta\beta'}.
\label{kg18}
\end{equation}
Finally, we write down the completeness relation for these wave functions,
\begin{equation}
\frac{1}{2mc}\sum_{\beta}\int\frac{\dd^3{\bm p}}{(2\pi)^3}\,\ii\beta\Phi_{\bm p}^{(\beta)}(x)\overleftrightarrow{\partial_0}[\Phi_{{\bm p}'}^{(\beta')}(x)]^*
=\delta^{(3)}({\bm x}-{\bm x}').
\label{kg19}
\end{equation}
Since the one-particle solutions of the Klein-Gordon equation~\eqref{kg9} form a complete and orthonormal set of functions (see, also Refs.~\cite{Grib1,Grib2}), 
one can use them to construct the boson field operator~\eqref{bf1}.

\subsubsection{Bogolyubov transformation for the boson field}
\label{bosons}

The instantaneous Hamiltonian of the bosonic field in the presence of an external time-dependent electric field is given by~\cite{Greiner1}
\begin{align}
\op{H}(t)=\frac{1}{2m}\int\dd^3{\bm x}\,\Bigl[\frac{1}{c^2}\dot{\op{\Phi}}^\dagger(x)\dot{\op{\Phi}}(x)
+\op{\Phi}^\dagger(x)\Bigl((\op{\bm p}-e{\bm A}(t))^2+(mc)^2\Bigr)\op{\Phi}(x)\Bigr].
\label{bf3}
\end{align}
Substituting here~\eqref{bf1}, we arrive at
\begin{align}
\op{H}(t)=\int\frac{\dd^3{\bm p}}{(2\pi)^3}\Bigl[\gamma^{(++)}_{\bm p}(t)\op{b}_{\bm p}^\dagger\op{b}_{\bm p}+
\gamma^{(+-)}_{\bm p}(t)\op{b}_{\bm p}^\dagger\op{d}_{-{\bm p}}^\dagger
+\gamma^{(-+)}_{\bm p}(t)\op{d}_{-{\bm p}}\op{b}_{\bm p}+\gamma^{(--)}_{\bm p}(t)\op{d}_{-{\bm p}}\op{d}_{-{\bm p}}^\dagger\Bigr],
\label{bf4}
\end{align}
where the time-dependent coefficients $\gamma_{\bm p}^{(\beta\beta')}(t)$ are defined as
\begin{equation}
\gamma_{\bm p}^{(\beta\beta')}(t)=\left\{ 
\begin{array}{ll}
\displaystyle\frac{1}{2mc^2}\Bigl(|\dot{\Phi}_{\bm p}^{(\beta)}(t)|^2+\omega_{\bm p}^2(t)|\Phi_{\bm p}^{(\beta)}(t)|^2\Bigr) &\quad{\mbox{if}}\,\,\beta=\beta',\\
\displaystyle\frac{1}{2mc^2}\Bigl([\dot{\Phi}_{\bm p}^{(\beta)}(t)]^*\dot{\Phi}_{\bm p}^{(\beta')}(t)+\omega_{\bm p}^2(t)
[\Phi_{\bm p}^{(\beta)}(t)]^*\Phi_{\bm p}^{(\beta')}(t)\Bigr)&\quad{\mbox{if}}\,\,\beta\neq\beta'.
\end{array}
\right.
\label{bf5}
\end{equation}
One can show using the asymptotic condition~\eqref{kg8} that 
\begin{equation}
\lim_{t\rightarrow-\infty}\gamma_{\bm p}^{(\beta\beta')}(t)=\omega_{\bm p}\delta_{\beta\beta'}.
\label{bf7}
\end{equation}
Thus, in the remote past the Hamiltonian~\eqref{bf4} describes a free bosonic field. It is the interaction with the pulsed electric field which leads to the
appearance of nondiagonal terms, $\op{b}_{\bm p}^\dagger\op{d}_{-{\bm p}}^\dagger$ and $\op{d}_{-{\bm p}}\op{b}_{\bm p}$, in~\eqref{bf4}. These terms, however, can be removed
by means of the Bogolyubov transformation~\cite{Bogolyubov}.

In order to diagonalize the Hamiltonian~\eqref{bf4}, we introduce new annihilation operators~\cite{Bogolyubov},
\begin{align}
\op{b}_{\bm p}(t)&=\eta_{\bm p}(t)\op{b}_{\bm p}+\xi_{\bm p}(t)\op{d}^\dagger_{-{\bm p}},\label{bf8}\\
\op{d}_{\bm p}(t)&=\eta_{-{\bm p}}(t)\op{d}_{\bm p}+\xi_{-{\bm p}}(t)\op{b}_{-{\bm p}}^\dagger,
\label{bf9}
\end{align}
and the corresponding creation operators as well. They are defined through unknown time-dependent functions $\eta_{\bm p}(t)$ and $\xi_{\bm p}(t)$. 
As the instantaneous operators, $\op{b}_{\bm p}(t)$ and $\op{d}_{\bm p}(t)$, should evolve from the corresponding in-operators, $\op{b}_{\bm p}$ and $\op{d}_{\bm p}$,
we infer that
\begin{equation}
\lim_{t\rightarrow -\infty}\eta_{\bm p}(t)=1,\quad\lim_{t\rightarrow-\infty}\xi_{\bm p}(t)=0.
\label{bf10}
\end{equation}
In addition, after imposing the bosonic commutation relations on the new set of annihilation and creation operators, we obtain that at all times
\begin{equation}
|\eta_{\bm p}(t)|^2-|\xi_{\bm p}(t)|^2=1.
\label{bf11}
\end{equation}
Keeping this in mind, we rewrite the bosonic field operator~\eqref{bf1} as
\begin{equation}
\op{\Phi}(x)=\int\frac{\dd^3{\bm p}}{(2\pi)^3}\Bigl(\phi_{\bm p}^{(+)}(x)\hat{b}_{\bm p}(t)+\phi_{\bm p}^{(-)}(x)\hat{d}_{-{\bm p}}^\dagger(t)\Bigr),
\label{bf18}
\end{equation}
where the new functions
\begin{align}
\phi_{\bm p}^{(+)}(x)&=\eta_{\bm p}^*(t)\Phi_{\bm p}^{(+)}(x)-\xi_{\bm p}^*(t)\Phi_{\bm p}^{(-)}(x),\label{bf19}\\
\phi_{\bm p}^{(-)}(x)&=\eta_{\bm p}(t)\Phi_{\bm p}^{(-)}(x)-\xi_{\bm p}(t)\Phi_{\bm p}^{(+)}(x),\label{bf20}
\end{align}
have been introduced in compliance with the Bogolyubov transformation [Eqs.~\eqref{bf8} and~\eqref{bf9}]. Moreover, we assume that
\begin{equation}
\phi_{\bm p}^{(\beta)}(x)=\ee^{\ii {\bm p}\cdot {\bm x}-\ii\beta\int^t\dd t'\omega_{\bm p}(t')}\tilde{\phi}_{\bm p}^{(\beta)}(t).
\label{bf21}
\end{equation}
Thus, it follows from Eqs.~\eqref{kg2},~\eqref{bf19},~\eqref{bf20}, and~\eqref{bf21} that
\begin{align}
\Phi_{\bm p}^{(+)}(t)=\eta_{\bm p}(t)&\ee^{-\ii\int^t\dd t'\omega_{\bm p}(t')}\tilde{\phi}_{\bm p}^{(+)}(t)\nonumber\\
&+\xi_{\bm p}^*(t)\ee^{\ii\int^t\dd t'\omega_{\bm p}(t')}\tilde{\phi}_{\bm p}^{(-)}(t),\label{bf22}\\
\Phi_{\bm p}^{(-)}(t)=\eta_{\bm p}^*(t)&\ee^{\ii\int^t\dd t'\omega_{\bm p}(t')}\tilde{\phi}_{\bm p}^{(-)}(t)\nonumber\\
&+\xi_{\bm p}(t)\ee^{-\ii\int^t\dd t'\omega_{\bm p}(t')}\tilde{\phi}_{\bm p}^{(+)}(t).\label{bf23}
\end{align}
One can show that these functions solve Eq.~\eqref{kg3} provided that
\begin{equation}
\tilde{\phi}_{\bm p}^{(\beta)}(t)=\sqrt{\frac{mc^2}{\omega_{\bm p}(t)}},
\label{bf24}
\end{equation}
with the coefficients $\eta_{\bm p}(t)$ and $\xi_{\bm p}(t)$ coupled through the equations,
\begin{align}
\dot{\eta}_{\bm p}(t)&=\frac{\dot{\omega}_{\bm p}(t)}{2\omega_{\bm p}(t)}\,\xi_{\bm p}^*(t)\ee^{2\ii\int^t\dd t'\omega_{\bm p}(t')},\label{bf25}\\
\dot{\xi}_{\bm p}^*(t)&=\frac{\dot{\omega}_{\bm p}(t)}{2\omega_{\bm p}(t)}\,\eta_{\bm p}(t)\ee^{-2\ii\int^t\dd t'\omega_{\bm p}(t')}.\label{bf26}
\end{align}
This system of equations has to be solved with the initial conditions such that at time $t_0$, which is before the pulsed electric field
starts to act, $\eta_{\bm p}(t_0)=1$ and $\xi_{\bm p}(t_0)=0$. It is useful to introduce new coefficients,
\begin{align}
c_{\bm p}^{(1)}(t)&=\eta_{\bm p}(t)\ee^{-\ii\int^t\dd t'\omega_{\bm p}(t')},\label{bf27}\\
c_{\bm p}^{(2)}(t)&=\xi_{\bm p}^*(t)\ee^{\ii\int^t\dd t'\omega_{\bm p}(t')},\label{bf28}
\end{align}
as it allows to remove the rapidly oscillating in time phase factors in~\eqref{bf25} and~\eqref{bf26}. In this case, Eqs.~\eqref{bf25}
and~\eqref{bf26} become
\begin{equation}
\ii\frac{\dd}{\dd t}\begin{bmatrix}
                     c_{\bm p}^{(1)}(t)\\
                     c_{\bm p}^{(2)}(t)
                    \end{bmatrix}
                    =
                    \begin{pmatrix} \omega_{\bm p}(t) & -\ii\Omega_{\bm p}(t) \cr -\ii\Omega_{\bm p}(t) & -\omega_{\bm p}(t) \end{pmatrix}
                    \begin{bmatrix}
                     c_{\bm p}^{(1)}(t)\\
                     c_{\bm p}^{(2)}(t)
                    \end{bmatrix},
\label{bf29}
\end{equation}
where 
\begin{equation}
\Omega_{\bm p}(t)=-\frac{\dot{\omega}_{\bm p}(t)}{2\omega_{\bm p}(t)}=-ce{\cal E}(t)\frac{c(p_{\|}-eA(t))}{2\omega_{\bm p}^2(t)},
\label{bf30}
\end{equation}
and we impose the initial conditions such that $c_{\bm p}^{(1)}(t_0)=1$ and $c_{\bm p}^{(2)}(t_0)=0$.

In closing this section, let us rewrite the instantaneous Hamiltonian~\eqref{bf3} using the time-dependent operators. Namely, 
\begin{equation}
\op{H}(t)=\int\frac{\dd^3{\bm p}}{(2\pi)^3}\,\omega_{\bm p}(t)\Bigl(\op{b}_{\bm p}^\dagger(t)\op{b}_{\bm p}(t)+\op{d}_{-{\bm p}}^\dagger(t)\op{d}_{-{\bm p}}(t)\Bigr),
\label{bf17}
\end{equation}
where we have removed an infinite constant by normal ordering the operators $\op{d}_{-{\bm p}}(t)$ and $\op{d}_{-{\bm p}}^\dagger(t)$. As one can see, at each time $t$,
Eq.~\eqref{bf17} represents a collection of harmonic oscillators with energy $\omega_{\bm p}(t)$. Interestingly, the functions which define the Bogolyubov transformation
and, thus, allow to diagonalize the Hamiltonian, are obtained from solutions of~\eqref{bf29}. Hence, for a single eigenmode of the bosonic field, the problem is 
equivalent to solving a two-level system~\eqref{bf29} whose dynamics is not determined by a unitary matrix. The latter is an element of the $SU(1,1)$
group, the properties of which are analyzed in Appendix~\ref{SU(1,1)}. Further, we will investigate physical consequences of a nonunitary character of time evolution 
of the bosonic field as compared to the fermionic case.

\subsubsection{Normalized charge distribution of created boson pairs}
\label{mombos}

Before we define the quantity that will be analyzed in Sec.~\ref{numerical}, we go back to Eqs.~\eqref{bf8} and~\eqref{bf9}. These equations define an instantaneous vacuum state $\ket{0_t}$ such that $\op{b}_{\bm p}(t)\ket{0_t}=0$ 
and $\op{d}_{\bm p}(t)\ket{0_t}=0$. It is different than the in-vacuum state, as 
\begin{equation}
\op{b}_{\bm p}(t)\ket{0_{-\infty}}=\xi_{\bm p}(t)\op{d}_{-{\bm p}}^\dagger\ket{0_{-\infty}}, \quad
\op{d}_{\bm p}(t)\ket{0_{-\infty}}=\xi_{-{\bm p}}(t)\op{b}_{-{\bm p}}^\dagger\ket{0_{-\infty}}.
\label{extra}
\end{equation}
In addition, the charge field operator can be derived from~\eqref{kg13},
\begin{equation}
\op{Q}=e\int\frac{\dd^3{\bm p}}{(2\pi)^3}\Bigl(\op{b}_{\bm p}^\dagger(t)\op{b}_{\bm p}(t)-\op{d}_{-{\bm p}}^\dagger(t)\op{d}_{-{\bm p}}(t)\Bigr),
\label{charge-op}
\end{equation}
where the normal ordering of the creation and annihilation operators has been introduced. As it follows from~\eqref{extra}, the mean value of $\op{Q}$ in the in-vacuum 
state is zero. One can also show that the charge field operator is conserved during the time-evolution. Building upon the definition of~\eqref{charge-op}, we can interpret 
\begin{align}
Q^{(0)}({\bm p},t)&=e\bra{0_{-\infty}}\op{b}_{\bm p}^\dagger(t)\op{b}_{\bm p}(t)\ket{0_{-\infty}}\nonumber\\
&=e\bra{0_{-\infty}}\op{d}_{-{\bm p}}^\dagger(t)\op{d}_{-{\bm p}}(t)\ket{0_{-\infty}}\nonumber\\
&=e|\xi_{\bm p}(t)|^2=e|c_{\bm p}^{(2)}(t)|^2\label{bf12}
\end{align}
as the charge distribution of created bosons with momentum ${\bm p}$ and antibosons with momentum $-{\bm p}$ from
the initial vacuum state by the pulsed electric field. Here, we have used Eq.~\eqref{bf28}. While this accounts for quasiparticles, the 
charge distribution of a real boson pair is obtained from~\eqref{bf12} by taking the limit $t\rightarrow +\infty$.
Moreover, when considered as a function of ${\bm p}$, it will be related to the momentum distribution of created bosons. We will
refer to it in Sec.~\ref{results}.

\subsection{Spinor QED}
\label{FF}

The spinor QED formulation of the pair production from vacuum by a time-dependent pulsed electric field has been presented in Ref.~\cite{KTK}
(see, also Refs.~\cite{Grib1,Grib2}). It is based on the Dirac equation that describes spin 1/2 particles, when coupled to an external electromagnetic field. 
As it was shown there, the dynamics of a single eigenmode of fermionic field, specified by the momentum ${\bm p}$
and the spin projection $\lambda$, is defined by two differential equations,
\begin{equation}
\ii\frac{\dd}{\dd t}\begin{bmatrix}
                     c_{\bm p}^{(1)}(t)\\
                     c_{\bm p}^{(2)}(t)
                    \end{bmatrix}
                    =
                    \begin{pmatrix} \omega_{\bm p}(t) & \ii\Omega_{\bm p}(t) \cr -\ii\Omega_{\bm p}(t) & -\omega_{\bm p}(t) \end{pmatrix}
                    \begin{bmatrix}
                     c_{\bm p}^{(1)}(t)\\
                     c_{\bm p}^{(2)}(t)
                    \end{bmatrix},
\label{a40}
\end{equation}
that are analogous to Eq.~\eqref{bf29}. This similarity is due to the fact that, for a linearly polarized electric field, the bi-spinor part of the fermionic wave function decouples~\cite{KTK,Grib1,Grib2}.
Hence, each eigenmode exhibits the same time-evolution irrespectively of the particles spins $\lambda$. This does not hold for a circularly or an elliptically polarized fields.
In these cases, the Dirac-Heisenberg-Wigner approach~\cite{IBB1} and its development based on the spinoral decomposition~\cite{IBB2} can be used instead. Going back to Eq.~\eqref{a40}, 
it has to be solved with the same initial conditions as in Sec.~\ref{BB}, $c_{\bm p}^{(1)}(t_0)=1$ and $c_{\bm p}^{(2)}(t_0)=0$. This time, however,
\begin{equation}
\displaystyle\Omega_{\bm p}(t)=-\frac{ce{\cal E}(t)\epsilon_\perp}{2\omega_{\bm p}^2(t)},
\label{wrzuta5}
\end{equation}
where $\epsilon_\perp=\sqrt{(c{\bm p}_\perp)^2+(mc^2)^2}$. Another difference is that while the matrix governing the time evolution here is Hermitian, 
for the bosonic case it is pseudo-Hermitian [see, Eq.~\eqref{bf29}]. This, in principle, may have far-reaching consequences which will be studied in detail next.

In closing this section, we note that the charge distribution of created fermion pairs with momenta ${\bm p}$ and $-{\bm p}$ for a particle and an anti-particle, respectively, is
\begin{equation}
Q^{(1/2)}({\bm p},t)=e|c_{\bm p}^{(2)}(t)|^2,\label{a28}
\end{equation}
where the coefficient $c_{\bm p}^{(2)}(t)$ satisfies~\eqref{a40}.

\section{Momentum distributions of created particles}
\label{results}

In this section, we shall analyze statistical effects in the electron-positron pair creation from vacuum under the influence of time-dependent, linearly polarized electric field pulses.
For this purpose, we will use the formulations introduced in Secs.~\ref{BB} and~\ref{FF}, treating the pairs as scalar or spinor particles.

We define the momentum distribution of created particles by a sequence of $N_{\rm rep}$ identical electric field pulses,
\begin{equation}
{\cal P}_{N_{\rm rep}}^{(s)}({\bm p})=\lim_{t\rightarrow +\infty} |c_{\bm p}^{(2)}(t)|^2,
\label{res1}
\end{equation}
which follows from Eqs.~\eqref{bf12} and~\eqref{a28}. Depending on the statistics, which is reflected in a different set of equations being solved
for $c_{\bm p}^{(2)}(t)$ [Eq.~\eqref{bf29} for spinless particles ($s=0$) and Eq.~\eqref{a40} for spinor particles $(s=1/2)$], we may observe different patterns 
in the momentum distributions~\eqref{res1}. For instance, it was shown in Ref.~\cite{Li1} that the longitudinal spectra of created bosons and fermions
are shifted by $\pi/2$. In relation to those results, we will focus here on the longitudinal spectra as well, i.e., we set ${\bm p}_\perp={\bm 0}$. 
However, before presenting our numerical results we shall derive the Fraunhofer-type formulas for pair creation from vacuum that arise in the scalar and spinor QED.

\subsection{Fraunhofer-type formulas for the scalar and spinor QED}
\label{Fraunhoferform}

Consider the time-evolution matrix $\op{U}(t,t_0)$ for a single eigenmode 
of either the bosonic or fermionic field. In accordance with Eqs.~\eqref{bf29} and~\eqref{a40}, it satisfies the equation
\begin{equation}
\ii\frac{\dd}{\dd t}\op{U}(t,t_0)
=\begin{pmatrix} \omega_{\bm p}(t) & \mp\ii\Omega_{\bm p}(t) \cr -\ii\Omega_{\bm p}(t) & -\omega_{\bm p}(t) \end{pmatrix}
\op{U}(t,t_0),
\label{res2}
\end{equation}
where the upper sign relates to the boson and the lower one to the fermion statistics. Note that $\Omega_{\bm p}(t)$ differs in both cases too [Eqs.~\eqref{bf30} and~\eqref{wrzuta5}]. 
It follows from~\eqref{res2} that for fermions the time evolution is unitary, meaning that $\op{U}$ belongs 
to the $SU(2)$ group (see, Appendix~\ref{appendix1}). However, for bosons this is not the case. One can prove using Eq.~\eqref{res2} that for bosons,
\begin{equation}
\frac{\dd}{\dd t}\bigl[\op{U}^\ddag(t,t_0)\op{U}(t,t_0)\bigr]=0,
\label{res3}
\end{equation}
where the pseudo-Hermitian conjugate of $\op{U}(t,t_0)$ has been introduced,
\begin{equation}
\op{U}^\ddag(t,t_0)=\op{\sigma}_3\op{U}^\dagger(t,t_0)\op{\sigma}_3,
\label{re333}
\end{equation}
with $\op{\sigma}_3=\begin{pmatrix} 1 & 0 \cr 0 & -1 \end{pmatrix}$. Hence, accounting for the initial condition $\op{U}(t_0,t_0)=\op{I}$, we obtain
\begin{equation}
\op{U}^\ddag(t,t_0)\op{U}(t,t_0)=\op{I}.
\label{res4}
\end{equation}
This means that $\op{U}$ is the element of the $SU(1,1)$ group, discussed in Appendix~\ref{SU(1,1)}. Keeping this in mind, we shall derive now physical consequences 
of unitary vs. pseudo-unitary time evolution of particles created from the vacuum by a sequence of electric field pulses.

\subsubsection{Monodromy matrix}
\label{mono}

For a train of $N_{\rm rep}$ identical electric field pulses, each of time duration $T$, both functions $\Omega_{\bm p}(t)$ and $\omega_{\bm p}(t)$ 
in~\eqref{res2} are periodic within the time interval $N_{\rm rep}T$, with a period $T$. 
Hence, the same applies to $\op{U}(t,t_0)$. This property combined with the composition condition for the time-evolution operators results in
\begin{equation}
\op{U}(N_{\rm rep}T+t_0,t_0)=\prod_{j=0}^{N_{\rm rep}-1}\op{U}\bigl((j+1)T+t_0,jT+t_0\bigr)=\bigl[\op{U}(T+t_0,t_0)\bigr]^{N_{\rm rep}},
\label{composition}
\end{equation}
where we shall refer to $\op{U}(T+t_0,t_0)$ as the {\it monodromy matrix}~\cite{Yakubovich}. This matrix is evaluated at the period of the interaction with the external
electric field. As a consequence of~\eqref{composition}, it determines the system evolution under the influence of a finite sequence of well-separated electric field pulses.

The monodromy matrix in the fermionic case has been introduced in~\cite{KTK} (see also, Appendix~\ref{appendix1}).  It was shown there that it can be parametrized
using four real parameters such that $0\leqslant \vartheta_0,\vartheta,\beta < 2\pi$ and $0\leqslant \gamma \leqslant\pi$,
\begin{equation}
\op{U}(T+t_0,t_0)=\ee^{-\ii \vartheta_0}
\begin{pmatrix}
\cos\vartheta+\ii\sin\vartheta\cos\gamma & \ii\ee^{-\ii\beta}\sin\vartheta\sin\gamma
\cr
\ii\ee^{\ii\beta}\sin\vartheta\sin\gamma & \cos\vartheta-\ii\sin\vartheta\cos\gamma
\end{pmatrix}.
\label{monodromyFF}
\end{equation}
One can check that its eigenvalues are 
\begin{align}
\lambda_1&=\ee^{-\ii\vartheta_1}\quad{\rm with}\quad\vartheta_1=\vartheta_0-\vartheta,\nonumber\\
\lambda_2&=\ee^{-\ii\vartheta_2}\quad{\rm with}\quad\vartheta_2=\vartheta_0+\vartheta,
\label{eigenvaluesFF}
\end{align}
meaning that $|\lambda_{1,2}|=1$. As it was presented in Ref.~\cite{KTK}, the respective phases $\vartheta_1,\vartheta_2\in\mathbb{R}$ play a significant role in interpreting interference 
patterns in the momentum distributions of created fermions. The same is true for bosons. The difference, however, is that for bosons the phases $\vartheta_1$ and $\vartheta_2$ are not necessarily real.

As shown in Appendix~\ref{SU(1,1)}, the pseudo-unitary monodromy matrix for bosons can be either parametrized as
\begin{align}
\op{U}(T+t_0,t_0)=\ee^{-\ii \vartheta_0}
\begin{pmatrix}
\cos\vartheta+\ii\sin\vartheta\cosh\gamma  &-\ii\ee^{-\ii\beta}\sin\vartheta\sinh\gamma
\cr
\ii\ee^{\ii\beta}\sin\vartheta\sinh\gamma  &\cos\vartheta-\ii\sin\vartheta\cosh\gamma
\end{pmatrix} ,
\label{monodromyBB}
\end{align}
with $0\leqslant \vartheta_0,\vartheta,\beta < 2\pi$ and $\gamma\geqslant 0$, or as
\begin{align}
\op{U}(T+t_0,t_0)=\ee^{-\ii \vartheta_0}
\begin{pmatrix}
\cosh\vartheta+\ii\sinh\vartheta\sinh\gamma  &-\ii\ee^{-\ii\beta}\sinh\vartheta\cosh\gamma
\cr
\ii\ee^{\ii\beta}\sinh\vartheta\cosh\gamma  &\cosh\vartheta-\ii\sinh\vartheta\sinh\gamma
\end{pmatrix} ,
\label{monodromyBBpseudo}
\end{align}
with $0\leqslant \vartheta_0,\beta < 2\pi$ and $\vartheta,\gamma\geqslant 0$. Both these matrices satisfy the condition~\eqref{res4} but their eigenvalues have different character. 
While in the first case the eigenvalues are given by~\eqref{eigenvaluesFF}, in the second case one finds that
\begin{align}
\lambda_1&=\ee^{-\ii\vartheta_1}\quad{\rm with}\quad\vartheta_1=\vartheta_0-\ii\vartheta,\nonumber\\
\lambda_2&=\ee^{-\ii\vartheta_2}\quad{\rm with}\quad\vartheta_2=\vartheta_0+\ii\vartheta,
\label{eigenvaluesBB}
\end{align}
where the phases $\vartheta_1,\vartheta_2\in\mathbb{C}$. In either case we have $\vartheta_1+\vartheta_2=2\vartheta_0$ and $|\lambda_1\lambda_2|=1$.

In closing, we note that all of the aforementioned parametrized matrices share the same property. Namely, if we denote $\op{U}(T+t_0,t_0)\equiv\op{U}(\vartheta_0,\vartheta;\beta,\gamma)$ then
it follows from explicit derivations that
\begin{equation}
\op{U}(\vartheta_0^{(1)}+\vartheta_0^{(2)},\vartheta^{(1)}+\vartheta^{(2)};\beta,\gamma)=\op{U}(\vartheta_0^{(1)},\vartheta^{(1)};\beta,\gamma)\op{U}(\vartheta_0^{(2)},\vartheta^{(2)};\beta,\gamma).
\label{sram1}
\end{equation}
This becomes important in light of Eq.~\eqref{composition}. Based on this property, one can show that (for details, see Appendices~\ref{SU(1,1)} and~\ref{appendix1})
\begin{equation}
\op{U}(N_{\rm rep}T+t_0,t_0)=[\op{U}(\vartheta_0,\vartheta;\beta,\gamma)]^{N_{\rm rep}}=\op{U}(N_{\rm rep}\vartheta_0,N_{\rm rep}\vartheta;\beta,\gamma).
\label{sram2}
\end{equation}
This has been already proven in Ref.~\cite{KTK} for the fermionic case. We have also realized there that the parameters $\beta$ and $\gamma$ do not play a role
in interpreting the interference patterns in momentum distributions of created pairs. Actually, the parameter $\vartheta_0$ does not either, as it only enters the formulas
through the global phase factor. This means that, up to an irrelevant value $\vartheta_0$, the phases of the eigenvalues of the monodromy
matrices considered in this paper can be chosen either real [Eq.~\eqref{eigenvaluesFF}] or purely complex [Eq.~\eqref{eigenvaluesBB}]. We will use this convention 
when presenting our numerical results. For completeness, let us note that $\vartheta_0$ depends on arbitrarily chosen phases of amplitudes $c_{\bm p}^{(1)}(t)$
and $c_{\bm p}^{(2)}(t)$ in the remote past.

\subsubsection{Diffraction and interference terms}
\label{terms}

In compliance with our current approach, the momentum distribution of created particles is defined as
\begin{equation}
{\cal P}_{N_{\rm rep}}^{(s)}=|\bra{-}\op{U}(+\infty,-\infty)\ket{+}|^2,
\label{res8}
\end{equation}
with the asymptotic in- and out-states, $\ket{+}=\begin{pmatrix} 1\\0 \end{pmatrix}$ and $\ket{-}=\begin{pmatrix} 0\\1 \end{pmatrix}$, corresponding to the free particle and antiparticle
energy, $+\omega_{\bm p}$ and $-\omega_{\bm p}$ respectively. Note that beyond the time interval $(N_{\rm rep}T+t_0,t_0)$ the fields evolve freely. For this reason, Eq.~\eqref{res8} reduces to
\begin{equation}
{\cal P}_{N_{\rm rep}}^{(s)}=|\bra{-}\op{U}(N_{\rm rep}T+t_0,t_0)\ket{+}|^2.
\label{res9}
\end{equation}
Using here Eq.~\eqref{sram2} with appropriately chosen parametrizations of matrices $\op{U}$ [Eqs.~\eqref{monodromyFF},~\eqref{monodromyBB}, or~\eqref{monodromyBBpseudo}], we obtain that
the momentum distribution of pairs created from vacuum by a sequence of $N_{\rm rep}$ electric field pulses equals
\begin{align}
{\cal P}_{N_{\rm rep}}^{(s)}&=\left\{  \begin{array}{ll} \sinh^2\gamma\sin^2(N_{\rm rep}\vartheta) & \mbox{for}\,\, s=0\;\mbox{and}\;\vartheta_1,\vartheta_2\in\mathbb{R},\\
				   \cosh^2\gamma\sinh^2(N_{\rm rep}\vartheta) & \mbox{for}\,\, s=0\;\mbox{and}\;\vartheta_1,\vartheta_2\in\mathbb{C},\\
                                   \sin^2\gamma\sin^2(N_{\rm rep}\vartheta)  & \mbox{for}\,\, s=1/2,
                        \end{array}
       \right. \label{res10}
\end{align}
while for an individual pulse,
\begin{align}
{\cal P}_1^{(s)}&=\left\{  \begin{array}{ll} \sinh^2\gamma\sin^2\vartheta  & \mbox{for}\,\, s=0\;\mbox{and}\;\vartheta_1,\vartheta_2\in\mathbb{R},\\
				   \cosh^2\gamma\sinh^2\vartheta & \mbox{for}\,\, s=0\;\mbox{and}\;\vartheta_1,\vartheta_2\in\mathbb{C},\\
                                   \sin^2\gamma\sin^2\vartheta  & \mbox{for}\,\, s=1/2.
                        \end{array}
       \right. \label{res12}
\end{align}
Hence, we obtain a standard Fraunhofer-type formula~\eqref{intro1}, i.e.,
\begin{equation}
{\cal P}_{N_{\rm rep}}^{(s)}={\cal P}_1^{(s)}\Bigl[\frac{\sin(N_{\rm rep}\vartheta)}{\sin\vartheta}\Bigr]^2,
\label{fraun1}
\end{equation}
which is valid for fermions~\cite{KTK} and for bosons provided that $\vartheta_1,\vartheta_2\in\mathbb{R}$. Here, we recognize that ${\cal P}_1^{(s)}$ plays a role of the diffraction
term, while $\bigl[\frac{\sin(N_{\rm rep}\vartheta)}{\sin\vartheta}\bigr]^2$ is a typical interference term. In contrast, a new type of Fraunhofer formula arises for bosons in the case
when $\vartheta_1,\vartheta_2\in\mathbb{C}$. Namely,
\begin{equation}
{\cal P}_{N_{\rm rep}}^{(s)}={\cal P}_1^{(s)}\Bigl[\frac{\sinh(N_{\rm rep}\vartheta)}{\sinh\vartheta}\Bigr]^2,
\label{fraun2}
\end{equation}
which also can be obtained from~\eqref{fraun1} by replacing $\vartheta$ by ${\rm i}\vartheta$. Therefore, by analogy with~\eqref{fraun1}, we shall still interpret
${\cal P}_1^{(s)}$ as a diffraction whereas $\bigl[\frac{\sinh(N_{\rm rep}\vartheta)}{\sinh\vartheta}\bigr]^2$ as an interference term. Irrespectively of the case considered, the latter
depends only on the parameter $\vartheta$ and the number of pulses in a train. This, in turn, relates to the eigenvalues of the monodromy matrix such that $\vartheta_2-\vartheta_1=2\vartheta$ in Eq.~\eqref{fraun1}
or $\vartheta_2-\vartheta_1=2\ii\vartheta$ in Eq.~\eqref{fraun2}.

Let us first discuss the case when $\vartheta_2-\vartheta_1=2\vartheta$. It follows from~\eqref{res12} that whenever this phase difference is zero (modulo $2\pi$),
which happens for $\vartheta=n\pi$ where $n=0,\pm 1,\pm 2,...$, the momentum distribution ${\cal P}_1^{(s)}$ vanishes. This is definitely the case of fermions. 
For bosons, however, it happens provided that $\sinh^2\gamma$ is not simultaneously infinite.
The same conclusion can be drawn from Eq.~\eqref{res10}
for ${\cal P}_{N_{\rm rep}}^{(s)}$, even though in this case additional zeroes occur. Now, consider $\bar{\vartheta}=n\pi+\delta\vartheta$ where $\delta\vartheta\ll 1$.
This means that there is a small phase difference between both eigenvalues of the monodromy matrix, $\vartheta_2-\vartheta_1=2\delta\vartheta$ (modulo $2\pi$), known as the {\it avoided crossing}~\cite{KTK,KKproc}.
One can check that for $\bar{\vartheta}$ the interference term in Eq.~\eqref{fraun1} behaves like
\begin{equation}
\Bigl[\frac{\sin(N_{\rm rep}\vartheta)}{\sin\vartheta}\Bigr]^2\Biggl|_{\vartheta=\bar{\vartheta}}\approx N_{\rm rep}^2\Bigl[1-\frac{1}{3}(N_{\rm rep}^2-1)(\delta\vartheta)^2\Bigr].
\label{res15}
\end{equation}
Hence, for as long as
\begin{equation}
|\delta\vartheta|\ll\sqrt{\frac{3}{N_{\rm rep}^2-1}},
\label{res16}
\end{equation}
one should observe a nearly perfect coherent enhancement of momentum distributions of produced pairs. Note that perfectly coherent enhancement, i.e., characterized by the scaling factor $N_{\rm rep}^2$,
can never be reached. The reason being that, even though the interference term scales like $N_{\rm rep}^2$ when $\vartheta=n\pi$, as shown by our numerical examples,
at those points ${\cal P}_1^{(s)}$ and ${\cal P}_{N_{\rm rep}}^{(s)}$
are both zero. Moreover, it follows from the general theory presented in this section that the momentum distributions calculated for a single pulse~\eqref{res12} should be more regular that the ones induced 
by a train of pulses~\eqref{res10}. This can be inferred from the fact that in-between every two subsequent zeroes of ${\cal P}_1^{(s)}$ there is additional $(N_{\rm rep}-1)$ zeroes of 
${\cal P}_{N_{\rm rep}}^{(s)}$. They occur for such parameters for which $\vartheta=n\pi+\frac{m\pi}{N_{\rm rep}}$, where $m=1,2,...,(N_{\rm rep}-1)$. Also, in-between two subsequent zeroes
of ${\cal P}_1^{(s)}$ there is $(N_{\rm rep}-2)$ local maxima of the momentum distributions at
$\vartheta=n\pi+\frac{(2m+1)\pi}{2N_{\rm rep}}$, where $m=1,2,...,(N_{\rm rep}-2)$. These are actually minor maxima, observed for $N_{\rm rep}>2$. Most importantly, the spectra exhibit major maxima 
for $N_{\rm rep}>1$, which scale nearly like $N_{\rm rep}^2$ [Eq.~\eqref{res15}]. Note that the aforementioned properties concern the fermionic and bosonic pair production, 
with some restrictions imposed on the latter. Namely, this is provided that the eigenvalues of the corresponding monodromy matrix~\eqref{monodromyBB} 
have real phases.

Going to the case $\vartheta_2-\vartheta_1=2\ii\vartheta$, one concludes from Eqs.~\eqref{res10} and~\eqref{res12} that the probability distributions ${\cal P}_{N_{\rm rep}}^{(0)}$
and ${\cal P}_1^{(0)}$ would be zero at $\vartheta=0$, provided that $\cosh^2\gamma$ is not infinite. As follows from our numerical examples, this is not the case.
Surprisingly, when the spectrum of the respective monodromy matrix~\eqref{monodromyBBpseudo} becomes degenerate 
($\vartheta_1=\vartheta_2$), the momentum distributions of created bosons do scale as $N_{\rm rep}^2$. We will refer to those points as exceptional points~\cite{Rotter1,Rotter2,Praxmeyer}. 
In the vicinity of such points, i.e., at the avoided crossings $\bar{\vartheta}=\delta\vartheta\ll 1$, the interference term in Eq.~\eqref{fraun2} becomes
\begin{equation}
\Bigl[\frac{\sinh(N_{\rm rep}\vartheta)}{\sinh\vartheta}\Bigr]^2\Biggl|_{\vartheta=\bar{\vartheta}}\approx N_{\rm rep}^2\Bigl[1+\frac{1}{3}(N_{\rm rep}^2-1)(\delta\vartheta)^2\Bigr].
\label{res15bis}
\end{equation}
Similarly, for this to occur, Eq.~\eqref{res16} has to hold. In contrast however to the previous case, ${\cal P}_{N_{\rm rep}}^{(0)}$
neither exhibits additional zeroes nor secondary maxima, compared to ${\cal P}_1^{(0)}$.

Note that the results presented in this section are not restricted to the process of pair creation. The reason being that our starting point was the set of differential equations~\eqref{res1}.
We would like to stress that these general equations describe the dynamics (either unitary or pseudo-unitary) of any two-level system exposed to a time-dependent, repetitive interaction. 
Therefore, our current predictions do apply to a variety of problems. Having said that, in the next section we will confront these predictions with the numerical results of momenta distributions of
pairs extracted from the vacuum by a finite sequence of identical electric field pulses.

\subsection{Numerical results}
\label{numerical}

\begin{figure}
\includegraphics[width=0.8\textwidth]{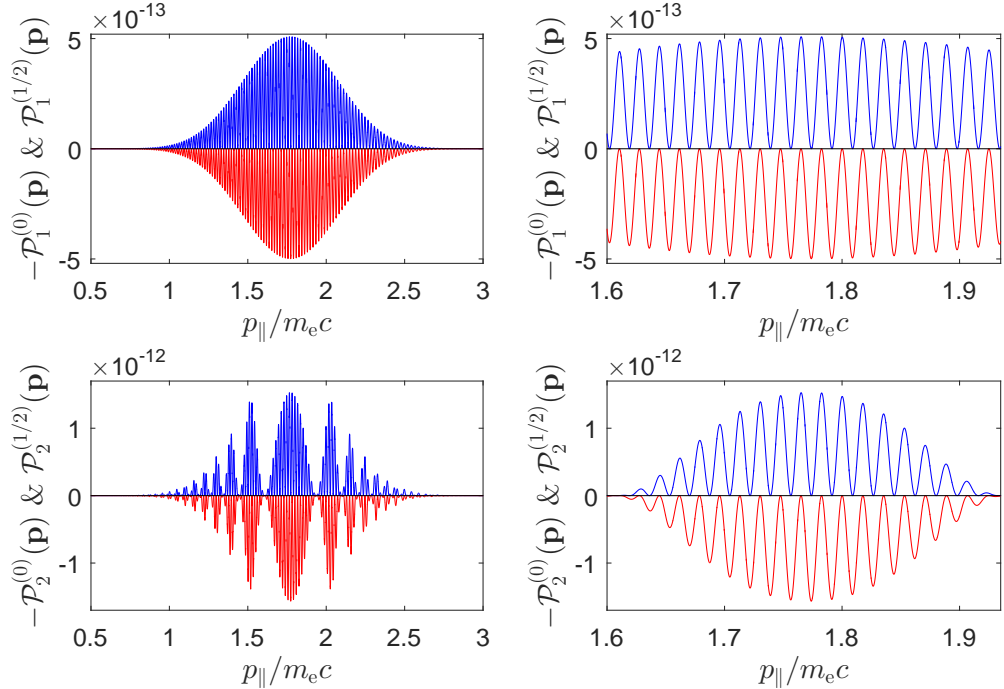}
\caption{Longitudinal momentum distributions of fermions (in blue) and bosons (in red) produced by a single ($N_{\rm rep}=1$) and a double ($N_{\rm rep}=2$) Gaussian pulse ($M=1$) with 
$\sigma=20/t_C$, $T_0=200t_C$, $T=400t_C$, and $\mathcal{E}_0=-0.1\mathcal{E}_S$. The distributions in the right column are the portions of the distributions from the left column. 
\label{dirackleinp02r12pair20190216}}
\end{figure}

In Fig.~\ref{dirackleinp02r12pair20190216}, we present the longitudinal momentum distributions~\eqref{res1} of created fermions (in blue) and the mirror-reflected 
distributions for bosons (in red). The spectra in the upper panels have been obtained for a single pulse ($N_{\rm rep}=1$), whereas the spectra in the lower panels
are for a sequence of two such pulses ($N_{\rm rep}=2$). Here, the Gaussian envelope ($M=1$) has been used, with the remaining field parameters being
$\sigma=20/t_C$, $T_0=200t_C$, $T=400t_C$, and $\mathcal{E}_0=-0.1\mathcal{E}_S$. As shown in Fig.~\ref{dirackleinpola}, in such case the two half-pulses are well separated
but, coincidentally, $T=2T_0$. Thus, the spectra of fermions and bosons are shifted by $\pi/2$.
Such a shift of the Fraunhofer-like peaks has been seen before and interpreted as originating from different statistics~\cite{Li1}. As we argue below, this is rather accidental and cannot be considered as statistical effect.

\begin{figure}
\includegraphics[width=0.9\textwidth]{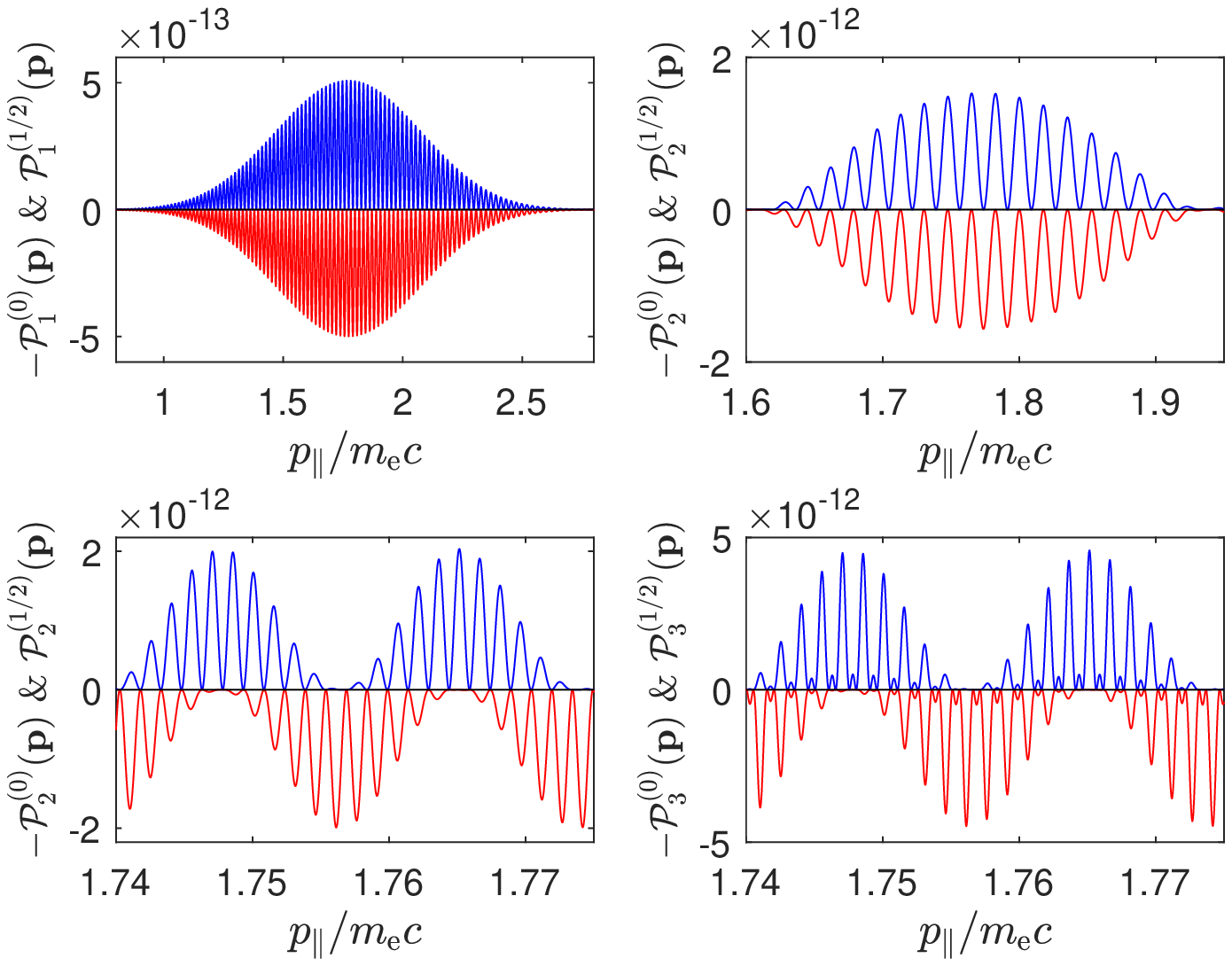}
\caption{For comparison, in the upper row we have reproduced the longitudinal momentum distributions of created fermion (blue) and boson (red) pairs from Fig.~\ref{dirackleinp02r12pair20190216}. In the left
panel, we show the spectra for $N_{\rm rep}=1$, whereas in the right panel for $N_{\rm rep}=2$. We recall that these results are for $T=2T_0$. The spectra for the same field parameters, except that
$T=14T_0$ are presented in the lower row. This time, the left panel is for $N_{\rm rep}=2$ and the right panel for $N_{\rm rep}=3$. 
\label{dkp02r123}}
\end{figure}

Based on the discussion in section~\ref{Fraunhoferform}, the pattern in the upper row of Fig.~\ref{dirackleinp02r12pair20190216} will be called the diffraction pattern.
This is to emphasize that it originates from interaction of the vacuum with a single electric field pulse. It is composed of rapid oscillations within 
a broad envelope. These diffraction peaks are such that the maxima of the distribution for fermions coincide with the zeroes of the distribution for bosons;
i.e., they are shifted by $\pi/2$. This, however, concerns the diffraction pattern and, as we have checked, is typical for a Gaussian pulse.
This is also corroborated by the previous results of Dumlu and Dunne~\cite{Dumlu1,Dumlu2} who have investigated the pair creation by a single 
Gaussian pulse with a carrier wave.

If the pairs are created by a train of two pulses, the diffraction pattern is multiplied by the interference term.
Thus, within the envelope ${\cal P}_1^{(s)}$, we should observe twice that dense series of peaks. Instead, in the lower row of Fig.~\ref{dirackleinp02r12pair20190216},
only an additional modulation of the spectra occurs. We shall show that this is accidental, as the extra interference peaks fall onto the zeroes of the diffraction
pattern. In addition, irrespectively of the statistics, the spectra exhibit a typical $N_{\rm rep}^2$-like scaling predicted by the Fraunhofer 
formula. 

\begin{figure}
\includegraphics[width=0.9\textwidth]{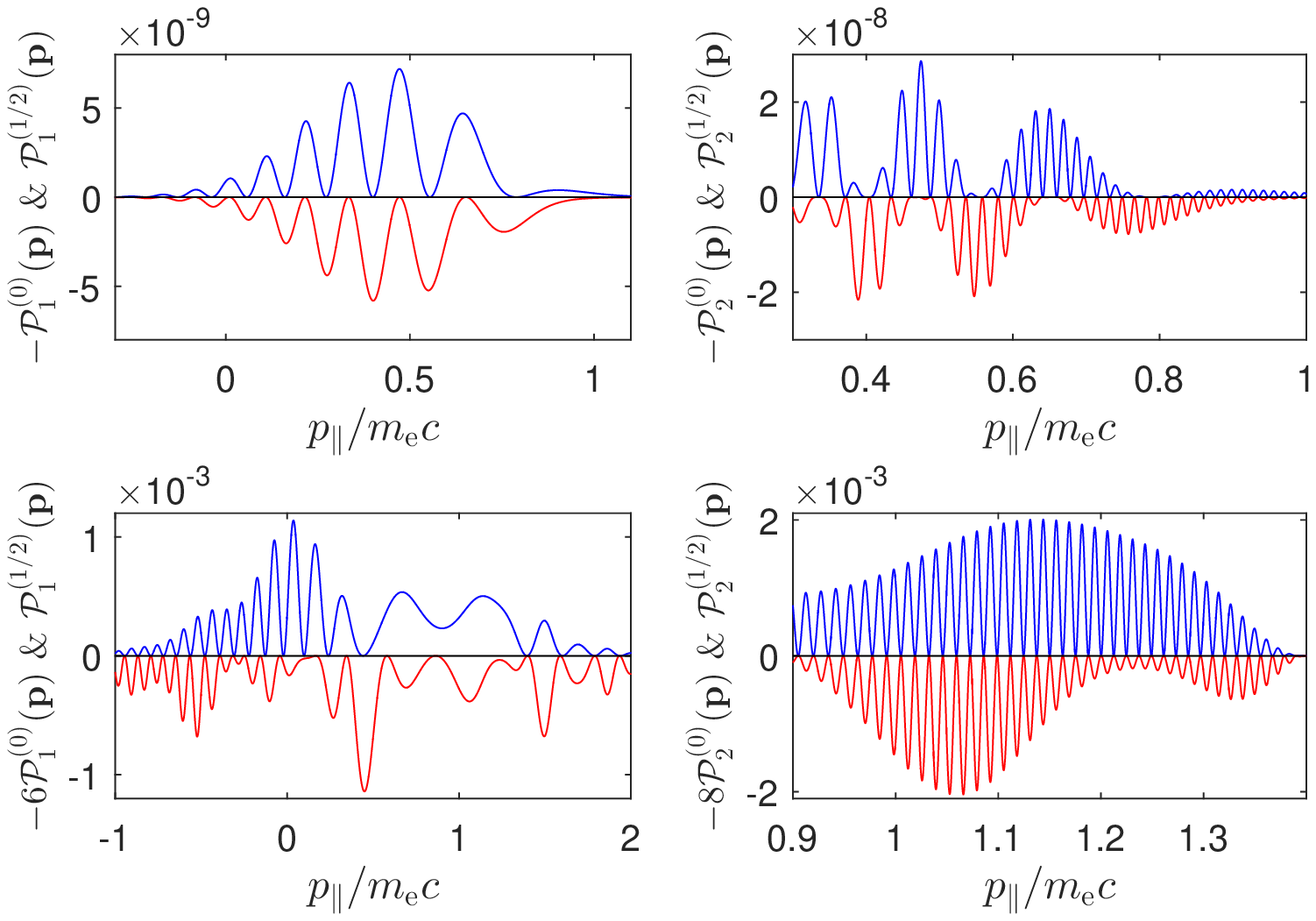}
\caption{Longitudinal momentum spectra  of fermions (blue) and bosons (red) for the external field parameters: $\sigma=5/t_C$, $T_0=40t_C$, $T=357t_C$, and $\mathcal{E}_0=-0.1\mathcal{E}_S$.
The results for Gaussian ($M=1$) and super-Gaussian ($M=5$) shaped pulses are shown in the upper and the lower rows, respectively. For $M=1$ we observe that the diffraction
maxima are shifted by $\pi/2$, whereas the interference maxima coincide with each other. Such general rule cannot be formulated for $M=5$. For instance, for the diffraction pattern
(bottom left panel) there are momentum regions ($p_{\|}<0.3m_\mathrm{e}c$ and $p_{\|}>1.2m_\mathrm{e}c$) where the peaks in the fermionic and bosonic spectra coincide, there is the region ($0<p_{\|}<0.3m_\mathrm{e}c$)
where they are shifted by roughly $\pi/2$, and the region $0.3m_\mathrm{e}c<p_{\|}<1.2m_\mathrm{e}c$ where both spectra exhibit slow modulations. On the other hand, both distributions $\mathcal{P}_2^{(s)}$
(bottom right panel) peak at the same values of $p_\|$. 
\label{dirackleinp02i10r12pair20190302}}
\end{figure}

To make our point, in Fig.~\ref{dkp02r123} we confront the spectra presented in Fig.~\ref{dirackleinp02r12pair20190216} with the spectra calculated
for the same field parameters, except that now $T=2800t_C$. In the upper row we show the diffraction peaks (left panel) and the modulated diffraction peaks (right panel)
for the case when either a single or a double pulse interacts with the vacuum, and $T=2T_0$. In the lower row, we show that within two diffraction peaks there is a fine 
peak structure, which originates from the interaction of either two (left panel) or three (right panel) pulses with the vacuum, with $T=14T_0$. 
Note that for $N_{\rm rep}=2$ this additional structure consists of major maxima, whereas for $N_{\rm rep}=3$ 
between two such maxima there appears a minor one. This is a typical interference pattern predicted by the Fraunhofer formula~\eqref{fraun1}. Most importantly, 
while the diffraction peaks are shifted by $\pi/2$, the interference peaks for fermions and bosons coincide. This clearly indicates 
that the respective shift of the bosonic and fermionic distributions is closely related to the parameters of the driving laser field, rather than to the statistics 
of created particles. We have confirmed this for other parameters as well. For instance, in Fig.~\ref{dirackleinp02i10r12pair20190302}
we present the longitudinal momentum distributions for a Gaussian (upper row) and a super-Gaussian envelope ($M=5$, lower row) for $\sigma=5/t_C$, $T_0=40t_C$, $T=357t_C$, 
$\mathcal{E}_0=-0.1\mathcal{E}_S$. One can see that sometimes the bosonic and fermionic spectra match very closely. This is the case of particles created from the vacuum 
by a sequence of two pulses (right column). For a single Gaussian pulse, there is $\pi/2$ shift between both momentum distributions. However, for a super-Gaussian pulse, there is a momentum region where both spectra 
coincide, but then they exhibit a shift which varies with the particle momentum. Interestingly, the results for a super-Gaussian envelope are by 5 orders of magnitude larger than 
for a Gaussian envelope (similar enhancement has been discussed in~\cite{Akall}). Since the shape effects are not the topic of this paper, they will be analyzed elsewhere. Here,
we shall look more closely into the peak structure of the presented momentum distributions.

\subsection{Interpretation of the results}
\label{interpretation}
\begin{figure}\
\includegraphics[width=0.9\textwidth]{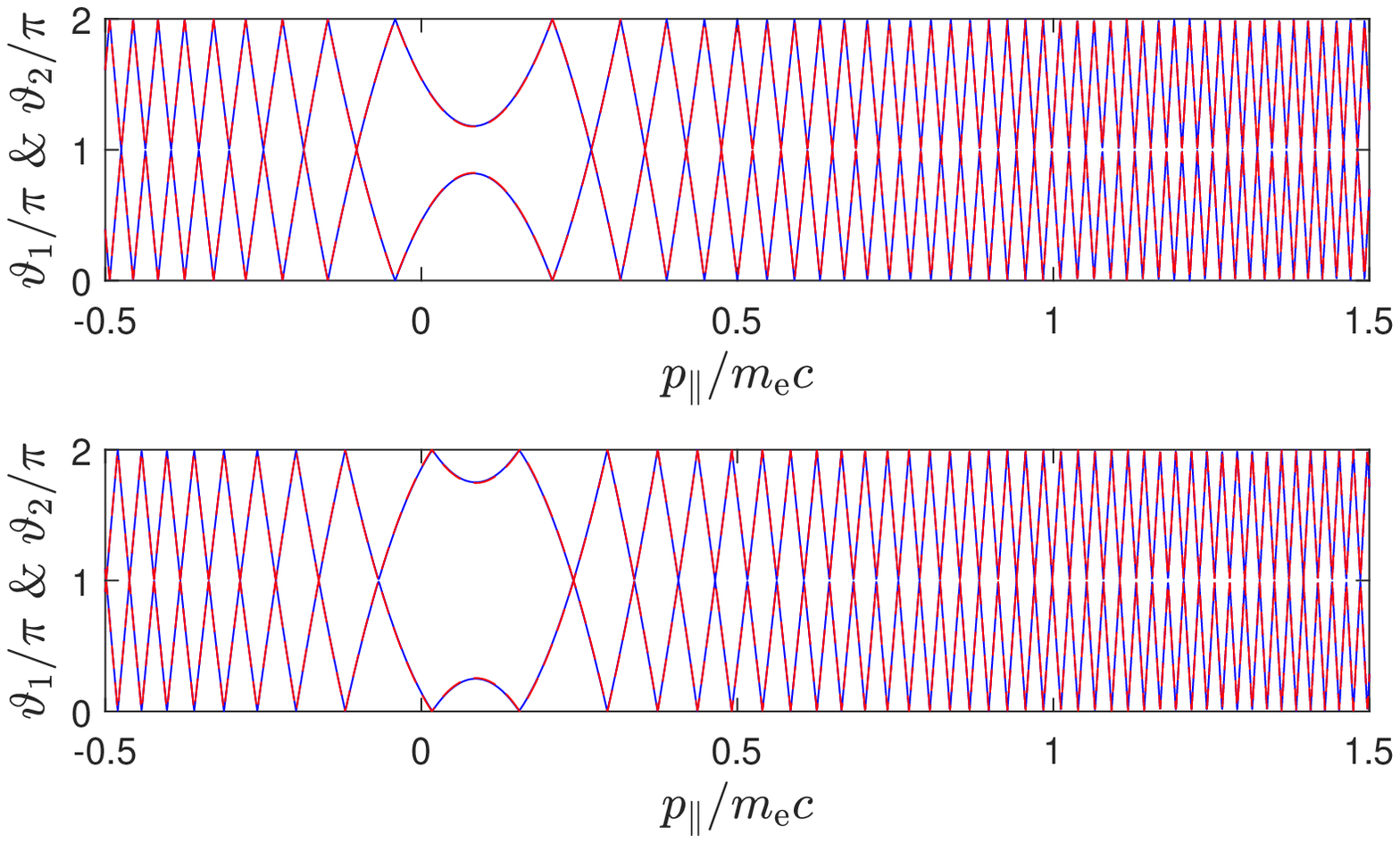}
\caption{The phases of eigenvalues of the monodromy matrix for $\sigma=5/t_C$, $T_0=40t_C$, $T=357t_C$, $\mathcal{E}_0=-0.1\mathcal{E}_S$, $N_{\mathrm{rep}}=1$ and $M=1$ 
(upper panel) or $M=5$ (lower panel). In each panel there are two lines: a solid blue line for spinor QED and the dashed red line for scalar QED. On the scale of the figure these lines are identical,
which occurs also for other parameters of the external field. In both cases and nearly for the same values of $p_\|$ we observe either true avoided crossings (for fermions) or 
pseudo-avoided crossings (for bosons). As it follows from the general theory formulated in section~\ref{Fraunhoferform},
both types of avoided crossings lead to coherent-like enhancements in the momentum distribution of created pairs.
\label{dirackleinsp02i10v2e01}}
\end{figure}

We have demonstrated numerically that the $N_{\rm rep}^2$-type enhancement of the momentum spectra of created particles is observed independently of their statistics. 
This is supported by Eqs.~\eqref{fraun1} and~\eqref{fraun2} and the analysis of phases $\vartheta_1$ and $\vartheta_2$ around the avoided crossings (see, Sec.~\ref{Fraunhoferform}). 
In Fig.~\ref{dirackleinsp02i10v2e01}, in both panels
we present these phases as functions of the longitudinal momentum of created bosons (dashed red curve) and fermions (solid blue curve). The upper panel is
for the Gaussian ($M=1$) whereas the lower panel is for the super-Gaussian ($M=5$) envelope. The remaining parameters are the same as in
Fig.~\ref{dirackleinp02i10r12pair20190302}, and $N_{\rm rep}=1$. It is clear that, on the scale of the figures, the red and blue curves are essentially the same. In other 
words, there is no obvious difference between the eigenvalues determining the time evolution of the bosonic and fermionic fields. This is quite surprising 
taking into account that there is a fundamental difference between both theories. Namely, while the time evolution for the fermionic field is unitary,
for the bosonic field it is pseudo-unitary (for more details, see Appendices~\ref{SU(1,1)} and~\ref{appendix1}). Since the difference between the respective
phases is nearly zero, the crossings and the avoided crossings in both cases occur at roughly same values of $p_\|$. This should result in a very similar 
interference pattern, which is confirmed in the right column of Fig.~\ref{dirackleinp02i10r12pair20190302}. In this column the so-called interference maxima and the zeroes of the distributions are
basically the same, even though the diffraction patterns with their own peaks and zeroes might differ.

\begin{figure}
\includegraphics[width=0.9\textwidth]{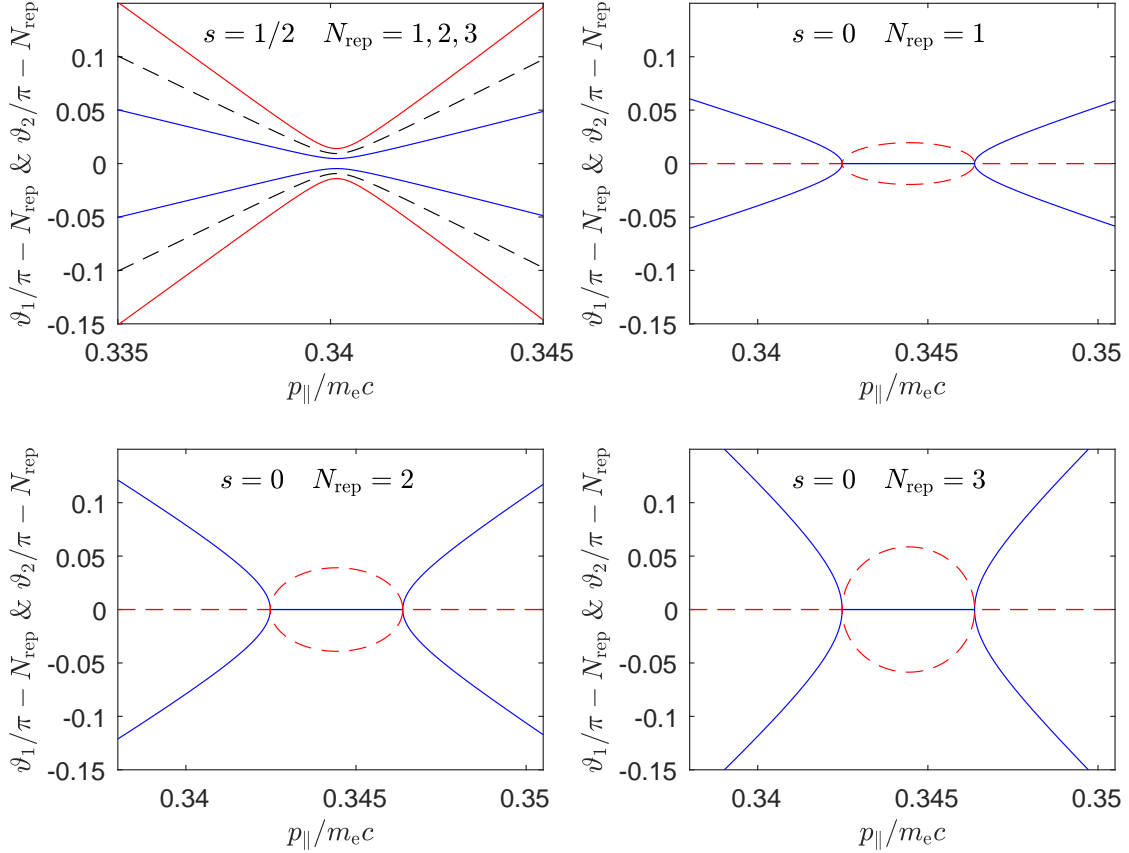}
\caption{The phases of the eigenvalues of monodromy matrices for fermions (top left panel) and bosons (remaining panels) in the case when the pairs are created by a train of $N_{\rm rep}$
electric super-Gaussian pulses ($M=5$) such that $\sigma=10/t_C$, $T_0=10t_C$, $T=200t_C$, and $\mathcal{E}_0=-0.5\mathcal{E}_S$. We show a true avoided crossing for fermions, 
with a non-zero gap between the phases which grows linearly with $N_{\rm rep}$. Pseudo-avoided crossings for bosons have very distinct features, as they occur in the vicinity of
exceptional points at which the phases $\vartheta_j$ ($j=1,2$) from being real turn to be
purely imaginary. Here, the real and imaginary parts of the phases are plotted as solid and dashed lines, respectively. Note that the phase difference in the complex regime also grows linearly with $N_{\rm rep}$.
\label{avoideddk190301}}
\end{figure}

An analysis of phases $\vartheta_1$ and $\vartheta_2$ for the case considered in Fig.~\ref{dirackleinp02r12pair20190216} predicts the same: The interference
pattern should look similar irrespectively of the statistics. At first glance, this is not the case in Fig.~\ref{dirackleinp02r12pair20190216}.
We have confronted this figure with the positions of avoided crossings. It turns out that every avoided crossing coincides interchangeably with 
the zero or the maximum of the diffraction pattern. For this reason, the spectra in Fig.~\ref{dirackleinp02r12pair20190216} are missing every other interference
peak, which invalidates the conclusion of Ref.~\cite{Li1}. As it follows from our analysis,
the positions of interference peaks are nearly the same for bosons and fermions. They are modulated, however, by the diffraction term that depends
on the particle statistics and on the parameters of the external field.

We have demonstrated that the positions of avoided crossings of the phases $\vartheta_1$ and $\vartheta_2$ are basically the same. In order to see a more pronounced difference
between both types of avoided crossings, in Fig.~\ref{avoideddk190301} we plot them for a stronger electric field and for a super-Gaussian pulse. More
specifically, these results have been obtained for $\sigma=10/t_C$, $T_0=10t_C$, $T=200 t_C$, ${\cal E}_0=-0.5{\cal E}_S$, and $M=5$. In this figure, 
the dependence of the phases on the longitudinal momentum of created particles around a chosen avoided crossing is presented for either fermions (top left panel)
or bosons (remaining panels). As denoted in each panel, the results illustrate the behavior of $\vartheta_1$ and $\vartheta_2$ for the cases when the pair
creation is stimulated by either a single ($N_{\rm rep}=1$), a double ($N_{\rm rep}=2$), or a triple ($N_{\rm rep}=3$) electric field pulse.
Note that here $\vartheta_0=\pi$, and so the total global phase accumulated over the entire field duration is $N_{\rm rep}\pi$.
For this reason, on the vertical axis we subtract $N_{\rm rep}$ from $\vartheta_{1,2}/\pi$.
In the fermionic case, we observe an actual avoided crossing, with a small gap that increases linearly with the number of pulses in the pulse sequence
(see, also Ref.~\cite{KTK}). The phases in the bosonic case reveal a different behavior around, what we call, a {\it pseudo-avoided crossing}. Around such crossing,
the phases $\vartheta_1$ and $\vartheta_2$ change from real (solid line) to complex (dashed line) values. This happens at the exceptional points.
In-between them, there is a single pseudo-avoided crossing where we observe a nearly perfect coherent enhancement in agreement with Eq.~\eqref{fraun2}.
Note also that, similar to the fermionic case, the corresponding gap increases linearly with $N_{\rm rep}$. In both cases it holds also that $\vartheta_1+\vartheta_2=0$ modulo $2\pi$.
Finally, we note that a peculiar behavior of phases in the bosonic case is observed in a very narrow momentum interval. Except of such intervals, the phases take real values
and their behavior is determined by~\eqref{fraun1}. Thus, while the major interference peaks can be attributed to pseudo-avoided crossings, the other features of the interference
pattern are the same as in the fermionic case. The same stays true for other field parameters as well, provided that the electric pulses are well separated, i.e., $T\gg T_0$.

\begin{figure}
\includegraphics[width=0.93\textwidth]{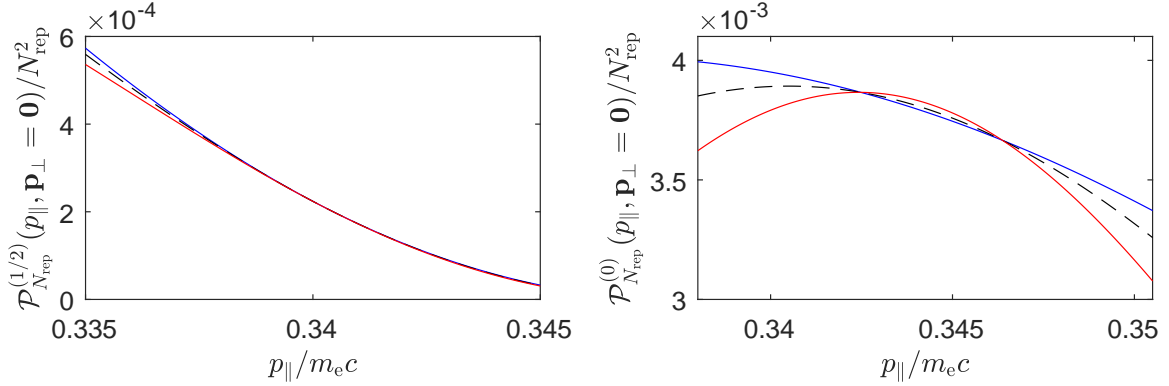}
\caption{The momentum distributions of created fermions (left panel) and bosons (right panel) near the avoided crossings shown in Fig.~\ref{avoideddk190301}.
The distributions are scaled by $N_{\rm rep}^2$. The blue line is for $N_{\rm rep}=1$, the dashed line for $N_{\rm rep}=2$, whereas the red line corresponds to $N_{\rm rep}=3$.
For fermions, the phase difference is small enough so the approximation~\eqref{res15} is well satisfied. For bosons, there are two momenta $p_\|$ for which all the distributions
take the same value. Rather than that, they follow either the approximation~\eqref{res15} or~\eqref{res15bis} depending on the momentum interval. As explained in the text,
at these two values of $p_\|$ the bosonic spectra are coherently enhanced. This is different than in the fermionic case, when the strictly coherent enhancement can never be observed.
\label{avoideddk0pair190301}}
\end{figure}

To demonstrate a very good agreement of our numerical results with the general theory presented in section~\ref{Fraunhoferform}, we plot
the momentum distributions around the avoided crossings analyzed in Fig.~\ref{avoideddk190301}. In Fig.~\ref{avoideddk0pair190301}, we show the corresponding 
distributions for fermions (left panel) and bosons (right panel) when the pairs are generated by various pulse configurations: $N_{\rm rep}=1$ (blue line),
$N_{\rm rep}=2$ (black line), and $N_{\rm rep}=3$ (red line). In the regions of $p_\|$ where the phases $\vartheta_j$ ($j=1,2$) are real, the curves decrease in magnitude
with increasing $N_{\rm rep}$. This is in agreement with the approximation originated from the standard Fraunhofer formula~\eqref{res15}, as the gap between both phases increases linearly with $N_{\rm rep}$
(with a small negative, qubic correction). 
On the other hand, in the momentum interval where the phases $\vartheta_j$ are complex, which is in a very small momentum interval for bosons,
we observe the opposite tendency. Here, the curves increase in magnitude with increasing $N_{\rm rep}$, as the phase difference $(\vartheta_2-\vartheta_1)$ in Eq.~\eqref{res15bis} 
increases linearly with $N_{\rm rep}$ (with a small positive, qubic correction).
Interestingly, for bosons there are two values of longitudinal momentum where all momentum distributions, when scaled by $N_{\rm rep}^2$, take the same value. This happens at exactly same
momenta for which $\vartheta_1=\vartheta_2$ in Fig.~\ref{avoideddk190301}, i.e., at the exceptional points. According to general theorem introduced in section~\ref{Fraunhoferform}, when the phase difference is strictly
zero ($\vartheta=0$) the respective momentum distributions should be zero as well. We observe, however, that at both crossings the spectra take nonzero values. The reason being that at these two points
the parameter $\gamma$ in~\eqref{res10} and~\eqref{res12} becomes infinitely large. As a consequence, a fully coherent enhancement of the bosonic spectra is observed at the exceptional points.
This feature distinguishes the bosonic and fermionic momentum distributions and, as such, can surely be regarded as a statistical effect. It also means that for fermions the nearly 
perfect coherent enhancement can be lost by changing the parameters of the electric field, such as the number of pulses in a train. For bosons, however, this is not the case.

\section{Conclusions}
\label{conclusions}

We have studied the creation of bosonic and fermionic pairs by a finite sequence of $N_{\rm rep}$ identical, time-dependent electric field pulses. In the case considered in this paper, i.e., when the 
pulses are linearly polarized, both problems simplify to solving a two-level system of equations for time-evolution of a single field eigenmode. The difference is
that the time-evolution matrix is either unitary for fermions or pseudo-unitary for bosons. This also affects the resulting momentum
distributions of particles, which has been studied in this paper in great detail.

In relation to our problem, we have formulated a general theory of a two-level system exposed to a periodic, but finite, time-dependent interaction. The distinction between unitary
and pseudo-unitary time evolution was made. In both cases, we have derived the Fraunhofer-type formulas describing the inter-level transitions. These formulas have been used then 
to interpret our numerical results of momentum distributions of created pairs.

In analogy to the standard Fraunhofer formula~\eqref{intro1}, the momentum distributions of produced particles exhibit the interference peaks which are modulated by the diffraction pattern.
Whereas the major interference peaks in the fermionic case does not follow strictly an $N_{\rm rep}^2$ scaling, it does happen for bosons. In both cases, this has been attributed 
to adiabatic transitions at avoided crossings of the phases defining the time-evolution matrix. The difference is the avoided crossing itself. For bosons, for instance, it is observed at the passage 
between real and imaginary phases. When this happens, the perfect $N_{\rm rep}^2$ enhancement of momentum distributions of created bosons is observed.

When analyzing the momentum distributions of created particles we have observed that, for the Gaussian pulse profile, the diffraction patterns for bosons and fermions are shifted
such that maxima of the former correspond to zeroes of the latter. This seems quite coincidental, as already for the super-Gaussian pulse profile it is not true. 
Regardless, the diffraction pattern does modulate the interference peaks and so it influences the overall structure of momentum distributions. Still the interference pattern
looks nearly identical for fermions and bosons.

\section*{Acknowledgements}

This work is supported by the National Science Centre (Poland) under Grant No. 2014/15/B/ST2/02203.

\appendix

\section{Quantum Vlasov equation}
\label{Vlasov}

In this appendix, we shall derive the quantum Vlasov equations for fermions and bosons~\cite{Schmidt}, which have been solved in the same context by various authors (see, 
e.g.~\cite{Akal,Li2}). First, we introduce new quantities,
\begin{align}
f({\bm p},t)&=|c_{\bm p}^{(2)}(t)|^2,\nonumber\\
u({\bm p},t)&=c_{\bm p}^{(1)}(t)[c_{\bm p}^{(2)}(t)]^*+[c_{\bm p}^{(1)}(t)]^*c_{\bm p}^{(2)}(t),\label{vla1}\\
v({\bm p},t)&=\ii \bigl(c_{\bm p}^{(1)}(t)[c_{\bm p}^{(2)}(t)]^*-[c_{\bm p}^{(1)}(t)]^*c_{\bm p}^{(2)}(t)\bigr),\nonumber
\end{align}
where the coefficients $c_{\bm p}^{(1)}(t)$ and $c_{\bm p}^{(2)}(t)$ are given by~\eqref{bf27} and~\eqref{bf28}.
Note that each of these functions is real and $f({\bm p},t)$ defines the temporal momentum distribution of created pairs 
in the given eigenmode of either the bosonic or fermionic field, 
\begin{equation}
f({\bm p},t)=|c_{\bm p}^{(2)}(t)|^2.\label{special}
\end{equation}
The latter follows from Eqs.~\eqref{bf12} and~\eqref{a28}, and the definition of the function $c_{\bm p}^{(2)}(t)$.
By comparison with Eq.~\eqref{res1}, one can also conclude that the momentum distributions of created particles considered in this paper can be obtained as
\begin{equation}
{\cal P}_{N_{\rm rep}}^{(s)}({\bm p})=\lim_{t\rightarrow\infty}f({\bm p},t).
\end{equation}
Calculating the time derivative of the quantities~\eqref{vla1} and using the system of equations~\eqref{bf29} for bosons whereas~\eqref{a40} for fermions, we obtain that
\begin{align}
\dot{f}({\bm p},t)&=\pm\Omega_{\bm p}(t)u({\bm p},t),\nonumber\\
\dot{u}({\bm p},t)&=-2\omega_{\bm p}(t)v({\bm p},t)\pm 2\Omega_{\bm p}(t)[1\pm 2f({\bm p},t)],\label{vla2}\\
\dot{v}({\bm p},t)&=2\omega_{\bm p}(t)u({\bm p},t)\nonumber,
\end{align}
where the upper sign relates to bosons and the lower sign to fermions. Here, we have also used the fact that $|c_{\bm p}^{(1)}(t)|^2=1\pm f({\bm p},t)$. 
Since at the initial time $t_0$ we had $c_{\bm p}^{(2)}(t_0)=0$,
the aforementioned system of equations has to be solved with the initial conditions: $f({\bm p},t_0)=0$, $u({\bm p},t_0)=0$, and $v({\bm p},t_0)=0$.

In doing so, we introduce a complex function $\zeta({\bm p},t)=u({\bm p},t)+\ii v({\bm p},t)$ and the following abbreviation, $R({\bm p},t)=2\Omega_{\bm p}(t)[1\pm 2f({\bm p},t)]$.
Then, the last two equations of~\eqref{vla2} can be written in the form of the first-order, inhomogeneous, linear differential equation
for the unknown function $\zeta({\bm p},t)$, 
\begin{equation}
\dot{\zeta}({\bm p},t)=2\ii\omega_{\bm p}(t)\zeta({\bm p},t)\pm R({\bm p},t).\label{vla3}
\end{equation}
Solving first the homogeneous equation,
\begin{equation}
\dot{\zeta}({\bm p},t)=2\ii\omega_{\bm p}(t)\zeta({\bm p},t),\label{vla4}
\end{equation}
we find out that
\begin{equation}
\zeta({\bm p},t)=C\exp\Bigl(2\ii\int_{t_0}^t\dd\tau\omega_{\bm p}(\tau)\Bigr),\label{vla5}
\end{equation}
where $C$ is the integration constant. Varying this constant, $C=C(t)$, and plugging
\begin{equation}
\zeta({\bm p},t)=C(t)\exp\Bigl(2\ii\int_{t_0}^t\dd\tau\omega_{\bm p}(\tau)\Bigr),\label{vla6}
\end{equation}
into Eq.~\eqref{vla3}, we arrive at 
\begin{equation}
\dot{C}(t)=\pm R(t)\exp\Bigl(-2\ii\int_{t_0}^t\dd\tau\omega_{\bm p}(\tau)\Bigr).\label{vla7}
\end{equation}
The solution of this equation with the initial condition that $C(t_0)=0$ is
\begin{equation}
C(t)=\pm\int_{t_0}^t\dd\tau R(\tau)\exp\Bigl(-2\ii\int_{t_0}^\tau\dd\sigma\omega_{\bm p}(\sigma)\Bigr).\label{vla8}
\end{equation}
Hence, combining~\eqref{vla6} with~\eqref{vla8}, we obtain that
\begin{equation}
\zeta({\bm p},t)=\pm\int_{t_0}^t\dd\tau R(\tau)\exp\Bigl(2\ii\int_{\tau}^t\dd\sigma\omega_{\bm p}(\sigma)\Bigr),\label{vla9}
\end{equation}
or, equivalently,
\begin{align}
u({\bm p},t)&=\pm\int_{t_0}^t\dd\tau R(\tau)\cos\Bigl(2\int_{\tau}^t\dd\sigma\omega_{\bm p}(\sigma)\Bigr),\label{vla10}\\
v({\bm p},t)&=\pm\int_{t_0}^t\dd\tau R(\tau)\sin\Bigl(2\int_{\tau}^t\dd\sigma\omega_{\bm p}(\sigma)\Bigr).\label{vla11}
\end{align}
Finally, the quantum Vlasov equation is obtained by substituting~\eqref{vla10} into the first equation of~\eqref{vla2},
\begin{align}
\dot{f}({\bm p},t)=2\Omega_{\bm p}(t)\int_{t_0}^t\dd\tau\Omega_{\bm p}(\tau)\bigl(1\pm 2f({\bm p},\tau)\bigr)\cos\Bigl(2\int_{\tau}^t\dd\sigma\omega_{\bm p}(\sigma)\Bigr),\label{vla12}
\end{align}
with $\omega_{\bm p}(t)$ and $\Omega_{\bm p}(t)$ defined in Sec.~\ref{BB} for bosons and in Sec.~\ref{FF} for fermions.

\section{$SU(1,1)$ group}
\label{SU(1,1)}

Elements of the $SU(1,1)$ group are $2\times 2$ complex matrices, $\op{U}$, which satisfy the following relation,
\begin{equation}
\op{U}^\ddag=\op{U}^{-1},
\label{su1}
\end{equation}
where the pseudo-Hermitian conjugate of $\op{U}$ has been defined in~\eqref{re333}. One can also prove that the eigenvalues $\lambda_j$ ($j=1,2$) of the matrix $\op{U}$ 
are such that $|\lambda_1\lambda_2|=1$.

Applying the above definition to the general matrix,
\begin{equation}
\op{U}=\begin{pmatrix} U_{11} & U_{12} \cr U_{21} & U_{22}\end{pmatrix},
\label{su2}
\end{equation}
we conclude that the matrix elements $U_{j\ell}$ ($j,\ell=1,2$) have to fulfill the conditions,
\begin{align}
|U_{11}|^2-|U_{21}|^2 & =1, \nonumber \\
|U_{22}|^2-|U_{12}|^2 & =1, \nonumber \\
U_{11}^*U_{12}-U_{21}^*U_{22} &=0.
\label{su3}
\end{align}
This actually means that the matrix $\op{U}$ can be uniquely defined by four real parameters. It turns out that one can introduce these parameters such that
\begin{equation}
\hat{U}= \ee^{-\ii\vartheta_0}
\begin{pmatrix}
\cos\vartheta+\ii\sin\vartheta\cosh\gamma & -\ii\ee^{-\ii\beta}\sin\vartheta\sinh\gamma
\cr
\ii\ee^{\ii\beta}\sin\vartheta\sinh\gamma & \cos\vartheta-\ii\sin\vartheta\cosh\gamma
\end{pmatrix},
\label{su4}
\end{equation}
where $0\leqslant \vartheta_0,\vartheta,\beta < 2\pi$ and $\gamma\geqslant 0$. Or, another possibility is
\begin{equation}
\hat{U}= \ee^{-\ii\vartheta_0}
\begin{pmatrix}
\cosh\vartheta+\ii\sinh\vartheta\sinh\gamma & -\ii\ee^{-\ii\beta}\sinh\vartheta\cosh\gamma
\cr
\ii\ee^{\ii\beta}\sinh\vartheta\cosh\gamma & \cosh\vartheta-\ii\sinh\vartheta\sinh\gamma
\end{pmatrix},
\label{su4bis}
\end{equation}
with $0\leqslant \vartheta_0,\beta < 2\pi$ and $\vartheta,\gamma\geqslant 0$. Transition between these two representations occurs through the exceptional point, at which 
$\vartheta=0$ and $\gamma=\infty$, with the substitutions: $\vartheta\rightarrow{\rm i}\vartheta, \cosh\gamma\rightarrow-{\rm i}\sinh\gamma$,
and $\sinh\gamma\rightarrow-{\rm i}\cosh\gamma$. Below, we discuss consequences of each parametrization.

\subsection{Real phases of the eigenvalues}
\label{case1}

For the matrix~\eqref{su4}, there are two complex eigenvalues, $\lambda_1=\ee^{-\ii(\vartheta_0-\vartheta)}$ and $\lambda_2=\ee^{-\ii(\vartheta_0+\vartheta)}$, 
with real phases $(\vartheta_0\mp\vartheta)$. The corresponding eigenvectors are
\begin{align}
\ket{1}&=\ee^{\ii\psi_1}\begin{pmatrix} \ee^{-\ii\beta/2}\cosh(\gamma/2) \cr
\ee^{\ii\beta/2}\sinh(\gamma/2) \end{pmatrix}, \nonumber \\
\ket{2}&=\ee^{\ii\psi_2}\begin{pmatrix} \ee^{-\ii\beta/2}\sinh(\gamma/2) \cr
\ee^{\ii\beta/2}\cosh(\gamma/2) \end{pmatrix} ,
\label{su5}
\end{align}
which are defined up to irrelevant phase factors, $0\leqslant \psi_1, \psi_2<2\pi$. In other words, it holds that $\op{U}\ket{j}=\lambda_j\ket{j}$, where $j=1,2$,
and $|\lambda_j|=1$ in agreement with the general statement that $|\lambda_1\lambda_2|=1$. The eigenstates~\eqref{su5} are linearly independent and orthonormal in the sense that
\begin{equation}
\sscal{\chi_1}{\chi_2}=\bra{\chi_1}\op{\sigma}_3\ket{\chi_2}
\label{su1new}
\end{equation}
defines the pseudo-scalar product for arbitrary complex column vectors, $\ket{\chi_j}$, $j=1,2$. Using this definition, one derives that
\begin{equation}
\sscal{1}{1}=1,\quad\sscal{2}{2}=-1,\quad\sscal{1}{2}=0=\sscal{2}{1}.
\label{su6}
\end{equation}
In compliance with Eq.~\eqref{su1new}, we construct the pseudo-projectors onto the given eigenstate $\ket{j}$ such that
\begin{equation}
\op{P}_1=\ket{1}\bra{1}\op{\sigma}_3, \quad \op{P}_2=-\ket{2}\bra{2}\op{\sigma}_3.
\label{su7}
\end{equation}
In our case, this gives
\begin{align}
\hat{P}_1&= \frac{1}{2}\begin{pmatrix}
1+\cosh\gamma & -\ee^{-\ii\beta}\sinh\gamma \cr \ee^{\ii\beta}\sinh\gamma & 1-\cosh\gamma
\end{pmatrix} ,
\nonumber \\
\hat{P}_2&= \frac{1}{2}\begin{pmatrix}
1-\cosh\gamma & \ee^{-\ii\beta}\sinh\gamma \cr -\ee^{\ii\beta}\sinh\gamma & 1+\cosh\gamma
\end{pmatrix} .
\label{su8}
\end{align}
With these definitions, we obtain
\begin{equation}
\op{P}_j^{\ddagger}=\op{P}_j, \quad\op{P}_j\op{P}_{\ell}=\op{P}_j\delta_{j\ell}, \quad\op{P}_1+\op{P}_2=\op{I}.
\label{su9}
\end{equation}
In addition, the spectral decomposition holds,
\begin{equation}
\op{U}=\lambda_1\op{P}_1+\lambda_2\op{P}_2.
\label{su10}
\end{equation}
At this point, it is important to stress that for $\vartheta=0$ the spectrum of $\op{U}$ is degenerate; namely, $\lambda_1=\lambda_2$. 
It follows from the above definitions that in this case the matrix $\op{U}$ becomes trivial, $\op{U}=\lambda_1\op{I}$.
We also note that, due to the properties of the pseudo-projection operators~\eqref{su9}, the matrix $\op{U}$ raised to the power $N$ is
\begin{equation}
\op{U}^N=\lambda_1^N\op{P}_1+\lambda_2^N\op{P}_2,
\label{su11}
\end{equation}
or, more generally, for any analytic function $f$ we have
\begin{equation}
f(\op{U})=f(\lambda_1)\op{P}_1+f(\lambda_2)\op{P}_2.
\label{su12}
\end{equation}
Eq.~\eqref{su11} will be of particular importance in Sec.~\ref{results}.

\subsection{Complex phases of the eigenvalues}
\label{case2}

For the matrix~\eqref{su4bis}, the corresponding eigenvalues are $\lambda_1=\ee^{-\ii(\vartheta_0+\ii\vartheta)}$ and $\lambda_2=\ee^{-\ii(\vartheta_0-\ii\vartheta)}$,
meaning that their phases $(\vartheta_0\mp\ii\vartheta)$ are complex. The peculiar feature of the matrix~\eqref{su4bis} is that the corresponding eigenvectors 
cannot be normalized in the sense of Eq.~\eqref{su1new}. Namely, their norm is zero. Nevertheless, one can introduce another system of the pseudo-projection operators
\begin{align}
\hat{P}_1&= \frac{1}{2}\begin{pmatrix}
1+\ii\sinh\gamma & -\ii\ee^{-\ii\beta}\cosh\gamma \cr \ii\ee^{\ii\beta}\cosh\gamma & 1-\ii\sinh\gamma
\end{pmatrix} ,
\nonumber \\
\hat{P}_2&= \frac{1}{2}\begin{pmatrix}
1-\ii\sinh\gamma & \ii\ee^{-\ii\beta}\cosh\gamma \cr -\ii\ee^{\ii\beta}\cosh\gamma & 1+\ii\sinh\gamma
\end{pmatrix},
\label{extra3}
\end{align}
such that
\begin{equation}
\op{P}_1^\ddag=\op{P}_2,\quad \op{P}_2^\ddag=\op{P}_1,\quad\op{P}_j\op{P}_{\ell}=\op{P}_j\delta_{j\ell}, \quad \op{P}_1+\op{P}_2=\op{I}.
\label{appa4newnew}
\end{equation}
This means that the spectral decomposition~\eqref{su10} and its consequences still hold. 
In closing we note that, similarly to the case considered in section~\ref{case1}, for $\vartheta=0$ the eigenvalues of the matrix~\eqref{su4bis} are degenerate 
$\lambda_1=\lambda_2=\ee^{-\ii\vartheta_0}$ and the matrix~\eqref{su4bis} becomes trivial. 

\section{$SU(2)$ group}
\label{appendix1}

Consider elements of the $SU(2)$ group which are $2\times 2$ complex matrices, $\op{U}$, satisfying the condition,
\begin{equation}
\op{U}^\dagger=\op{U}^{-1}.
\label{su2a}
\end{equation}
They preserve the standard scalar product, $\scal{\chi_1}{\chi_2}$,
for arbitrary two-component complex vectors $\ket{\chi_j}$ ($j=1,2$). In addition, one can show that $\det\op{U}=1$, resulting in the complex eigenvalues $\lambda_j$
such that $|\lambda_j|=1$.

Taking the most general form of a $2\times 2$ matrix~\eqref{su2} and using the definition~\eqref{su2a}, we find out that
\begin{align}
|U_{11}|^2+|U_{21}|^2 & =1, \nonumber \\
|U_{12}|^2+|U_{22}|^2 & =1, \nonumber \\
U_{11}^*U_{12}+U_{21}^*U_{22} &=0.
\label{appa2}
\end{align}
Similar to the case studied in Appendix~\ref{SU(1,1)}, we need four real parameters to determine $\op{U}$ in a unique way. For the purpose of this paper, we shall represent it as
\begin{equation}
\hat{U}= \ee^{-\ii\vartheta_0}
\begin{pmatrix}
\cos\vartheta+\ii\sin\vartheta\cos\gamma & \ii\ee^{-\ii\beta}\sin\vartheta\sin\gamma
\cr
-\ii\ee^{\ii\beta}\sin\vartheta\sin\gamma & \cos\vartheta-\ii\sin\vartheta\cos\gamma
\end{pmatrix} ,
\label{appa7}
\end{equation}
where $0\leqslant \vartheta_0,\vartheta,\beta<2\pi$ and $0\leqslant\gamma\leqslant\pi$. It turns out that the eigenvalues of the unitary matrix~\eqref{appa7} are the same as the 
one discussed in Appendix~\ref{SU(1,1)}, Eq.~\eqref{su4}; namely, $\lambda_1=\ee^{-\ii(\vartheta_0-\vartheta)}$ and $\lambda_2=\ee^{\ii(\vartheta_0+\vartheta)}$. The
corresponding eigenvectors however are different,
\begin{align}
\ket{1}&=\ee^{\ii\psi_1}\begin{pmatrix} \ee^{-\ii\beta/2}\cos(\gamma/2) \cr
\ee^{\ii\beta/2}\sin(\gamma/2) \end{pmatrix}, \nonumber \\
\ket{2}&=\ee^{\ii\psi_2}\begin{pmatrix} -\ee^{-\ii\beta/2}\sin(\gamma/2) \cr
\ee^{\ii\beta/2}\cos(\gamma/2) \end{pmatrix},
\label{appa3ap}
\end{align}
where the global phases $\psi_j$ can be chosen arbitrary and are irrelevant in our further analysis. With the standard definition of the projection operators,
\begin{equation}
\hat{P}_{j}=\ket{j}\bra{j} ,\quad j=1,2,
\label{appa4}
\end{equation}
one can verify that
\begin{equation}
\op{P}_j^{\dagger}=\op{P}_j,\quad \op{P}_j\op{P}_{\ell}=\op{P}_j\delta_{j\ell}, \quad \op{P}_1+\op{P}_2=\op{I}.
\label{appa4new}
\end{equation}
For completeness, we write down the explicit form of these operators,
\begin{align}
\hat{P}_1&= \frac{1}{2}\begin{pmatrix}
1+\cos\gamma & \ee^{-\ii\beta}\sin\gamma \cr \ee^{\ii\beta}\sin\gamma & 1-\cos\gamma
\end{pmatrix} ,
\nonumber \\
\hat{P}_2&= \frac{1}{2}\begin{pmatrix}
1-\cos\gamma & -\ee^{-\ii\beta}\sin\gamma \cr -\ee^{\ii\beta}\sin\gamma & 1+\cos\gamma
\end{pmatrix} .
\label{appa5}
\end{align}
Finally, one can check that with the current definitions the spectral decomposition~\eqref{su10} holds, as well as related to it Eqs.~\eqref{su11} and~\eqref{su12}.
Based on the same arguments as in Appendix~\ref{SU(1,1)}, we conclude that for $\vartheta=0$ the unitary matrix $\op{U}$ becomes trivial, $\op{U}=\lambda_1\op{I}$.


\begin{thebibliography}{99}

\bibitem{Crawford}
F. S. Crawford, {\it Waves} (McGraw-Hill, New York, 1968).

\bibitem{Rayleigh}
J. W. S. Rayleigh, {\it The Theory of Sound, Vol. I} (MacMillan, London, 1894).

\bibitem{Whitham}
G. B. Whitham, {\it Linear and Nonlinear Waves} (John Wiley \& Sons, New York, 1974).

\bibitem{BornWolf}
M. Born and E. Wolf, {\it Principles of Optics} (Pergamon, Oxford, 1968).

\bibitem{Fremont}
F. Fr\'emont, {\it Young-Type Interferences with Electrons} (Springer, Berlin, 2014).

\bibitem{VanHove}
M. A. Van Hove, W. H. Weinberg, and C.M. Chan, {\it Low-Energy Electron Diffraction} (Springer, Berlin, 1986).

\bibitem{Silverman1995}
P. M. Silverman, {\it More Then One Mistery: Explorations in Quantum Interference} (Springer, New York, 1995).

\bibitem{Silverman2008}
P. M. Silverman, {\it Quantum Superposition} (Springer, Berlin, 2008).

\bibitem{Deymier}
P. Deymier and K. Runge, {\it Sound Topology, Duality, Coherence and Wave-Mixing} (Springer, Cham, 2017).

\bibitem{Peierls1979}
R. Peierls, {\it Surprises in Theoretical Physics} (Princeton, Princeton, 1979).

\bibitem{BaronePredazzi}
V. Barone and E. Predazzi, {\it High-Energy Particle Diffraction} (Springer, Berlin, 2002).

\bibitem{Potylitsyn}
A. P. Potylitsyn, M. I. Ryazanov, M. N. Strikhanov, and A. A. Tishchenko, {\it Diffraction Radiation from Relativistic Particles} (Springer, Berlin, 2010).

\bibitem{SmithPurcell}
S. J. Smith and E. M. Purcell, Phys. Rev. {\bf 92}, 1069 (1953).

\bibitem{Williams2006}
G. P. Williams, Rep. Prog. Phys. {\bf 69}, 301 (2006).

\bibitem{K1}
K. Krajewska, M. Twardy, and J. Z. Kami\'nski, Phys. Rev. A {\bf 89}, 052123 (2014).

\bibitem{F1}
F. Cajiao V\'elez, J. Z. Kami\'nski, and K. Krajewska, Atoms {\bf 7}, 34 (2019).

\bibitem{K2}
K. Krajewska, F. Cajiao V\'elez, and J. Z. Kami\'nski, Phys. Rev. A {\bf 91}, 062106 (2015).

\bibitem{K3}
K. Krajewska and J. Z. Kami\'nski, Phys. Rev. A {\bf 90}, 052108 (2014).

\bibitem{K5}
F. Cajiao V\'elez, K. Krajewska, and J. Z. Kami\'nski, Phys. Rev. A {\bf 91}, 053417 (2015).

\bibitem{K6}
K. Krajewska and J. Z. Kami\'nski, Phys. Lett. A {\bf 380}, 1247 (2016).

\bibitem{Akkermans}
E. Akkermans and G. Dunne, Phys. Rev. Lett. {\bf 108}, 030401 (2012).

\bibitem{Li1}
Z. L. Li, D. Lu, and B. S. Xie, Phys. Rev. D {\bf 89}, 067701 (2014).

\bibitem{Li2}
Z. L. Li, D. Lu, B. S. Xie, L. B. Fu, J. Liu, and F. F. Schen, Phys. Rev. D {\bf 89}, 093011 (2014).

\bibitem{KTK}
J. Z. Kami\'nski, M. Twardy, and K. Krajewska, Phys. Rev. D {\bf 98}, 056009 (2018).

\bibitem{KKproc}
K. Krajewska, W. Gac, M. Twardy, and J. Z. Kami\'nski, Journal of Physics: Conf. Series {\bf 1206}, 012018 (2019).

\bibitem{Schmidt}
S. Schmidt, D. Blaschke, G. R\"opke, S. A. Smolyansky, A. V. Prozorkevich, and V. D. Toneev, Int. J. Mod. Phys. E {\bf 7}, 709 (1998).

\bibitem{KK}
K. Krajewska and J. Z. Kami\'nski, Phys. Rev. A {\bf 100}, 012104 (2019).

\bibitem{Akal}
I. Akal, S. Villalba-Ch\'avez, and C. M\"uller, Phys. Rev. D {\bf 90}, 113004 (2014).

\bibitem{Grib1}
A. A. Grib, S. G. Mamaev, and V. M. Mostepanenko, {\it Vacuum Quantum Effects in Strong External Fields} (Atomizdat, Moscow, 1988).

\bibitem{Grib2}
A. A. Grib, V. M. Mostepanenko, and V. M. Frolov, Teor. Mat. Fiz. {\bf 13}, 377 (1972).

\bibitem{Greiner1}
W. Greiner, {\it Relativistic Quantum Mechanics} (Springer, Berlin, 2000).

\bibitem{Greiner2}
W. Greiner and J. Reinhardt, {\it Quantum Electrodynamics} (Springer, Berlin, 2009).

\bibitem{Bogolyubov}
C. J. Pethick and H. Smith, {\it Bose-Einstein Condensation in Dilute Gases} (Cambridge, Cambridge, 2002).

\bibitem{IBB1}
I. Bia\l ynicki-Birula, P. G\'ornicki, and J. Rafelski, Phys. Rev. D {\bf 44}, 1825 (1991).

\bibitem{IBB2}
I. Bia\l ynicki-Birula and \L . Rudnicki, Phys. Rev. D {\bf 83}, 065020 (2011).

\bibitem{Wodkiewicz}
K. W\'odkiewicz and J. H. Eberly, ‎J. Opt. Soc. Am. B {\bf 2}, 458 (1985).

\bibitem{Makris}
K. G. Makris, R. El-Ganainy, D. N. Christodoulides, and Z. H. Musslimani, Phys. Rev. Lett. {\bf 100}, 103904 (2008).

\bibitem{Guo}
A. Guo, G. J. Salamo, D. Duchesne, R. Morandotti, M. Volatier-Ravat, V. Aimez, G. A. Siviloglou, and D. N. Christodoulides, Phys. Rev. Lett. {\bf 103}, 093902 (2009).

\bibitem{Ruter}
C. E. R\"uter, K. G. Makris, R. El-Ganainy, D. N. Christodoulides, M. Segev, and D. Kip, Nat. Phys. {\bf 6}, 192 (2010).

\bibitem{Lee}
Y.-C. Lee, J. Liu, Y.-L. Chuang, M.-H. Hsieh, and R.-K. Lee, Phys. Rev. A {\bf 92}, 053815 (2015).

\bibitem{Peng}
B. Peng, S. K. Ozdemir, S. Rotter, H. Yilmaz, M. Liertzer, F. Monifi, C. M. Bender, F. Nori, and L. Yang, Science {\bf 346}, 328 (2014).

\bibitem{Feng}
L. Feng, Z. J. Wong, R.-M. Ma, Y. Wang, and X. Zhang, Science {\bf 346}, 972 (2014).

\bibitem{Torosov}
B. T. Torosov and N. V. Vitanov, Phys. Rev. A {\bf 96}, 013845 (2017).

\bibitem{Kus}
R. Grimaudo, A. S. M. de Castro, M. Ku\'s, and A. Messina, Phys. Rev. A {\bf 98}, 033835 (2018).

\bibitem{Rotter1}
I. Rotter, J. Phys. A: Math. Theor. {\bf 42} 153001 (2009).

\bibitem{Moiseyev}
N. Moiseyev, {\it Non-Hermitian Quantum Mechanics} (Cambridge University Press, Cambridge, 2011).

\bibitem{Rotter2}
I. Rotter and J. P. Bird, Rep. Prog. Phys. {\bf 78} 114001 (2015).

\bibitem{Praxmeyer}
L. Praxmeyer, P. Yang, and R.-K. Lee, Phys. Rev. A {\bf 93}, 042122 (2016).

\bibitem{Yakubovich}
V. A. Yakubovich and V. M. Starzhinskii, {\it Linear Differential Equations with Periodic Coefficients} (John Wiley and Sons, New York, 1975).

\bibitem{Dumlu1}
C. K. Dumlu and G. V. Dunne, Phys. Rev. Lett. {\bf 104}, 250402 (2010).

\bibitem{Dumlu2}
C. K. Dumlu and G. V. Dunne, Phys. Rev. D {\bf 83}, 065028 (2011).

\bibitem{Akall}
I. Akal, arXiv:1712.05368v2 (2019).


\end{thebibliography}
\end{document}